\documentclass[12pt]{article}
\usepackage{amssymb}
\usepackage{amsmath}
\usepackage[mathscr]{euscript}

\input{epsf}

\def\beq{\begin{eqnarray}}
\def\eeq{\end{eqnarray}}

\def\half{{ 1 \over 2}}

\def\del{\partial}

\def\scdot{\! \cdot \!}

\def\bra{\langle}    \def\ket{\rangle}

\def\a1{a_1}

\def\pr{{\prime}}

       \def\sgn{\hbox{sgn}}

      \def\ma1{m_{\a1}}

\def\lnabla{\displaystyle\mathop{\nabla}^{\leftarrow}}
\def\rnabla{\displaystyle\mathop{\nabla}^{\rightarrow}}

\def\calA{{\cal A}}    \def\calN{{\cal N}}
    \def\calO{{\cal O}}

    \def\calR{{\cal R}}

\def\calI{{\cal I}}

    \def\calY{{\cal Y}}
\def\calM{{\cal M}}

\newcommand{\CG}[6]{(#1 \; #2 \;\, #3\; #4 \,|\, #5 \;#6)}

\begin{document}
\setcounter{figure}{0}
\setcounter{table}{0}
\renewcommand{\thetable}{\arabic{table}}
\renewcommand{\thefigure}{\arabic{figure}}

%==================================================================
\begin{center}
{\large \bf Electromagnetic transitions of excited baryons\\ in a
deformed 
oscillator quark model\\ }
%==================================================================
\vspace*{0.5cm}
 {Atsushi Hosaka\footnote{e-mail address: hosaka@la.numazu-ct.ac.jp} \\
Numazu College of Technology, Numazu, Shizuoka 410-8501, Japan}

\vspace*{0.5cm}
{Miho Takayama\footnote{e-mail address: takayama@rcnp.osaka-u.ac.jp}
 and Hiroshi Toki\footnote{e-mail address: toki@rcnp.osaka-u.ac.jp}\\
Research Center for Nuclear Physics (RCNP), Osaka University, \\
Ibaraki, Osaka 567-0047, Japan}
\end{center}
\vspace*{1cm}
%
%==================================================================
\abstract{
We study electromagnetic transitions 
of excited baryons in a deformed 
oscillator quark model, where
baryon excited states are described as rotational bands of 
deformed intrinsic states.    
We describe all necessary tools to compute transition amplitudes 
in multipole basis, which are then related to 
the commonly used helicity amplitudes.  
We pay a special attention on the sign of the amplitudes as well as 
their absolute values by computing the photon and pion couplings 
simultaneously.
We have found that the effect of deformation 
on the transition amplitudes is rather weak.  
The difficulty in reproducing the empirical amplitude of the Roper 
state is discussed.  
 }
%==================================================================
%

\newpage

%==================================================================
%SECTION{Introduction}
\setcounter{equation}{0}
\renewcommand{\theequation}{\arabic{section}-\arabic{equation}}
\section{Introduction}
%==================================================================
Excited baryons as well as their ground states provide precious 
information on the dynamics of low energy QCD.  
Not only their masses but also various transitions are 
particularly important for the study of their structure and 
interactions. Recent and future experiments planned at facilities such as 
TJNAL, COSY, CELCIUS and SPring8 are aiming at obtaining more detailed 
information on hadron structure~\cite{recentexp}.  

Based on conventional effective models of baryons, e.g.,  quark
models,  one can draw a simple but rather unified picture for
both ground and excited states~\cite{bhad_text}.  
There, constituent quarks are placed in a confining 
potential and are interacting through suitable residual interactions.  
As a consequence, we have an intuitive picture similar to
the atomic and nuclear physics; 
baryon states are described as single 
particle states of three valence quarks.  

An extensive work was performed by Isgur and Karl in the
non-relativistic (NR) quark model which has been 
applied to a variety of hadronic phenomena~\cite{isg}, including 
electromagnetic couplings of excited states~\cite{KI}.
Recently, the role of the Nambu-Goldstone bosons has been emphasized 
in the context of the chiral quark model~\cite{glozman}.  
In these models, important physics is dictated by valence quarks which 
are confined inside hadrons and interacting each other through 
residual interactions.  In the Isgur-Karl model, 
one gluon exchanges are introduced, 
while in the chiral quark model flavor dependent forces generated by  
Nambu-Goldstone boson exchanges are included.  
These residual interactions yield hyper fine splittings in baryon 
masses with configuration mixing.  
In general, however, the quality of these models 
crucially depends on the type of residual interactions with 
considerable amount of parameters.  

In such a situation, we still hope to find a better 
description for the entire baryonic system.  
In this respect, we have recently found a remarkable systematics in 
light flavor ($u, d$ and $s$) baryons~\cite{hotato,tatoho}.  
The systematics applies essentially to all SU(3) baryons
including almost 80 \% of the observed states nominated by the
Particle Data Group~\cite{PDG1,PDG2}.  
The resulting common structure of baryon spectra then look very much 
like a rotational band.  This in turn implies that excited baryons are likely 
to be deformed in space.  
In fact, the idea of deformed baryons is not
new~\cite{tokidd,bhaduri}.  
However, our finding has shown that the picture holds for more  
baryons than it was originally thought.  
Furthermore, it turns out that a very simple non-relativistic 
quark model with a deformed oscillator potential can explain this
aspect at once. This is indeed a remarkable fact since the model 
essentially contains only one parameter.  
We call the model the deformed oscillator quark (DOQ) model as 
described in detail in 
Refs.~\cite{hotato,tatoho} 

Naturally as a next step,
 an interesting question arises; how does the DOQ model
describe various transition processes?
This is the issue we would like to address in this paper.  
For this we study the electromagnetic transitions of excited 
baryons.  In principle, we can consider various transitions between both 
excited and ground states. Experimentally, however, only transitions from 
the ground state nucleon to excited states are observed, where
experimental amplitudes are compiled in the form of the helicity 
amplitudes $A_{1/2}$ and $A_{3/2}$~\cite{PDG2,walker,moorhouse,PDG3}.  

We compute transition amplitudes first in the multipole basis in analogy with 
nuclear transitions. One significant difference, however, between the baryon and 
nuclear transitions is that in the latter the long wave length 
approximation can be 
used as the typical energy scale is a few MeV, while in the baryonic 
case, the same approximation can not be used since the relevant 
transition energy is compatible with the inverse length of a typical baryon size.

The plan of this paper is as follows.  
In section 2, we introduce the DOQ model and briefly discuss
the mass spectrum of SU(3) baryons~\cite{hotato,tatoho}.  
Then we present wave functions for the deformed excited states 
which are necessary for the computation of transition amplitudes.  
In section 3, we study electromagnetic transitions and 
derive various matrix elements.   
We discuss carefully theoretical amplitudes that can be compared with 
empirical amplitudes including their signs.  
Appropriate treatment of the sign 
requires additional information of the pion coupling, which we also 
discuss.   
In section 4, we compare theoretical amplitudes 
with experimental data.  
After discussing the results in the naive DOQ model, we consider 
two important effects which are missing in 
%takayama..
%the original DOQ model~\cite{hotato,tatoho}.  
our previous study.
They are diagonalization of the non-orthogonal basis and relativistic 
effects.    
We pay special attention to the transition of the Roper resonance,
since this is the channel for which theoretical explanation is difficult.  
Final section 5 is devoted to the summary of the present work.

%==================================================================
%SECTION{Deformed Oscillator Quark Model}
\setcounter{equation}{0}
\renewcommand{\theequation}{\arabic{section}-\arabic{equation}}
\section{Deformed Oscillator Quark Model}
%==================================================================

In this section we briefly discuss the DOQ model for excited 
baryons
~\cite{bhaduri,bohr}, 
and present 
necessary ingredients to compute transition amplitudes.

%------------------------------------------------------------------
\subsection{Deformed intrinsic states and rotational spectra}
%------------------------------------------------------------------

Let us start with the Hamiltonian of the DOQ model:
\beq
H^{\rm DOQ} = \sum_{i=1}^{3} \left(
\frac{\vec{p}_i^2}{2m} 
+ \frac{m}{2} ( \omega_x^2 x_i^2 
+ \omega_y^2 y_i^2 + \omega_z^2 z_i^2 ) \right) \; .
\label{Hho}
\eeq
Here the index $i$ runs over three valence quarks. 
The masses of constituent quarks $m$ 
is taken about  300 MeV for \( u \), \( d \) and \( s \) quarks.
Practically, in the harmonic oscillator model, only relevant 
parameters are the oscillator parameters $\omega_{i}$ and the actual 
value of $m$ is not important.  

Applying the following coordinate transformation
\beq
\vec R &=& \frac{1}{\sqrt{3}} 
( \vec x_1 + \vec x_2 + \vec x_3 ) \; ,
\nonumber \\
\vec \rho &=& \frac{1}{\sqrt{2}} ( \vec x_1 - \vec x_2 ) \; ,
\label{cotrans}  \\
\vec \lambda &=& \frac{1}{\sqrt{6}} ( \vec x_1 + \vec x_2 
-2 \vec x_3 ) \; , \nonumber
\eeq
the center of mass coordinate can be removed and the resulting 
intrinsic hamiltonian is  given as 
\beq
H^{\rm int} = \frac{\vec p_\rho^2}{2m} 
+ \frac{m}{2} \left( \omega_x^2 \rho_x^2 
+ \omega_y^2 \rho_y^2 + \omega_z^2 \rho_z^2 \right) 
+ (\rho \to \lambda) \; .
\label{Hint}
\eeq
The eigenstates of the intrinsic hamiltonian (\ref{Hint}) are 
specified by a set of oscillator quantum numbers
$(n_x^\rho, n_y^\rho, n_z^\rho ; \; 
  n_x^\lambda, n_y^\lambda, n_z^\lambda )$  
and the corresponding eigenenergies are given by 
\beq
E^{\rm int} &=&  (n_x^\rho + {\textstyle \frac{1}{2}}) \omega_{x}
+ (n_y^\rho + {\textstyle \frac{1}{2}}) \omega_{y}
+ (n_z^\rho + {\textstyle \frac{1}{2}}) \omega_{z} 
+ (\rho \to \lambda ) \nonumber \\
&=&
(N_x +1) \omega_x + (N_y +1) \omega_y + (N_z +1) \omega_z \; ,
\label{Eint}
\eeq
where $N_x = n_x^\rho + n_x^\lambda$,  
$N_y = n_y^\rho + n_y^\lambda$,  
$N_z = n_z^\rho + n_z^\lambda$.  
Thus the intrinsic energy (\ref{Eint}) is regarded as a function 
of $N_x$, $N_y$, \( N_{z} \) and \( \omega\)'s.

For a given $(N_x, N_y, N_z)$, the energy minimization is performed 
with respect 
to the deformation $\delta \omega_x$, $\delta \omega_y$ and 
$\delta \omega_z$.  
Unless we have some additional conditions, the variation leads to the 
trivial result; $E^{\rm int} \to 0$, when 
$\omega_x = \omega_y = \omega_z \to 0$.  
In order to avoid this collapse of the system, 
we impose the volume conservation condition
$\omega_x  \omega_y  \omega_z = \omega^3 = {\rm const}$.  
The validity of such an assumption depends on the underlying dynamics 
of quark confinement.  
In fact, one may relax the volume conservation by, for example, 
adding a term like 
$B/( \omega _x\omega _y\omega _z)^p$ 
to the energy (\ref{Eint}).  
It turns out, however, that the result does not change very much, 
and therefore we simply adopt the condition 
of volume conservation.  

Now the minimum energy of (\ref{Eint}) is given by 
\beq
E^{\rm int} = 3(N_x+1)^{1/3} (N_y+1)^{1/3} (N_z+1)^{1/3} \omega \; ,
\label{Emin}
\eeq
when
\beq
\omega_x : \omega_y : \omega_z  = 
\frac{1}{N_x+1} : \frac{1}{N_y+1} :\frac{1}{N_z+1} \, .
\label{r_omega}
\eeq
The inverse relation of (\ref{r_omega}) is slightly convenient as it 
gives the ratio of the lengths of axes of the deformed state:
\beq
a_x : a_y : a_z = N_x + 1 :  N_y + 1 : N_z + 1 \; ,
\eeq
where $a_{x,y,z}$ are root mean square lengths of the $x, y, z$
directions.  
The ground state with $N = 0$ $( N_x = N_y = N_z = 0)$ yields 
the spherically symmetric intrinsic state just as in 
the conventional quark model.  
For excited states $N = 1$ ( $N_x = N_y = 0$, $N_z = 1$ when the z-axis 
is chosen as a symmetry axis), the intrinsic state deforms 
prolately with the ratio of short to long axes 1 : 2.  
The energy of this deformed state is $ 3.780\omega$ 
as compared to $ 4\omega$ of the spherical quark model.  
For excited states $N=2$, there are two cases. 
One is $N_x = N_y = 0$, $ N_z=2$ for the prolate deformation 
with the ratio of short to long axes 1 : 3.  
The energy is $ 4.327\omega$ as compared to $ 5\omega$ of 
the spherical state with 
$(n,l) = (0,2)$ or 
$(1,0)$.  
Another is $N_x = N_y = 1, N_z=0$ which yields the oblate deformation 
with the ratio of short to long axes 1 : 2, whose energy is  
$4.762 \omega$.  
In general, the prolate shape takes the minimum energy for a given 
$N = N_x + N_y + N_z$, and so, 
in the following discussions we consider only the 
prolate deformation with the symmetry axis chosen along the $z$
direction. 
We summarize in Table~\ref{tbl:Eint} physical quantities of 
prolately deformed states. 

\begin{table}[htbp]
   \begin{center}         
   \caption{Various quantities of prolately 
     deformed states.
     Dimensional quantities are scaled by an appropriate power of $\omega$.  
     \label{tbl:Eint}}
     \vspace*{0.5cm}
     % \centering
     \begin{tabular}{ c c c c c c | c c c c c c }
         \hline
         \multicolumn{6}{ c|}{Positive parity} &
         \multicolumn{6}{ c}{Negative parity} \\
        \hline
          $N$  & $d $
          & $E^{\rm int}$ & $E^{\rm spherical}$ & $1/2\mathcal{I}$ & $\bra l^2 \ket$
          &$N$  & $d $
          & $E^{\rm int}$ & $E^{\rm spherical}$ & $1/2\mathcal{I}$ & $\bra l^2 \ket$\\
        \hline
        0  & 1  & --  & 3 & -- & 0 
                     &      &   &   &  &  \\
%         \hline
        2  & 3 & 4.33 & 5 & 0.0721 & 8   
                  &  1  & 2  & 3.78 & 4 & 0.126 & 3  \\
%         \hline
        4  & 5  & 5.13 & 7 & 0.0329 & 24 
                  &  3  & 4  & 4.76 & 6 & 0.0467 & 15 \\
%         \hline
        6  & 7  & 5.74 & 9 & 0.0191 & 48 
                  &  5  & 6  & 5.45 & 8 & 0.0246 & 35  \\
%         \hline
        \hline  
     \end{tabular}
   \end{center}
\end{table}

\begin{table}[htbp]
   \begin{center}     
       \begin{minipage}{15cm}
           \caption{
           Excitation
           energies of the $N=2$ and $N=1$ rotational 
           bands.  Energy values in units of MeV are computed by using 
           $\omega$ = 644 MeV. 
           %takayama
           ~\cite{tatoho} \label{tbl:Erot} }
       \end{minipage}
       \vspace*{0.5cm}
       % \centering
       \begin{tabular}{ c c c c | c c c c  }
           \hline
           \multicolumn{4}{ c |}{$N=2$} &
           \multicolumn{4}{ c}{$N=1$} \\
           %         \hline
           $l$  & & \multicolumn{2}{c|}{$E(N=2,l)-E(N=0)$} &
           $l$  & & \multicolumn{2}{c}{$E(N=1,l)-E(N=0)$} \\
           %         \hline
           & & in $\omega$ & in MeV & & & in $\omega$ & in MeV \\
           \hline
           0 & & 0.750 & 483 & 1 & & 0.654 & 421 \\
           2 & & 1.183 & 752 & 3 & & 1.914 & 1230 \\
           4 & & 2.192 & 1410 & 5 & & 4.182 & 2690 \\
           %         \hline
           \hline  
       \end{tabular}
   \end{center}
\end{table}

Since the deformed states break rotational symmetry, it should be 
recovered for eigenstates of angular momentum $l$.  
This can be performed by the standard cranking method~\cite{bohr}.  
The key quantities are the moment of inertia $\calI$ and the angular 
momentum fluctuation $\bra l^2 \ket$ of the deformed state.  
In the DOQ  model, they can be computed analytically.  
For a prolately deformed state of $z$-axis symmetry 
($\omega_x = \omega_y \neq \omega_z$), 
they are given by 
%takayama.. insert (N)
\beq
\calI_{N}&=&  \left( \frac{N_z + 1}{\omega_z}
+ \frac{N_y + 1}{\omega_y} \right) = \frac{1}{(N+1)^{1/3}} 
\left(  N^2 + 2N + 2 \right) \frac{1}{\omega} \, , \\
\left\bra l^2 \right\ket_{N} &=& \left(
\left(\frac{\omega_x}{\omega_{z}}\right)^2 -1 \right) 
=
N(N+2) \, , 
\eeq 
The rotational band is then constructed for each $N$: 
%takayama.. \hbar -> 1
\beq
\label{Erot}
E_{N\,l} = E^{\rm int}_{N} - \frac{1}{2\calI_{N}} \left\bra 
l^2\right\ket_{N}
+ \frac{1}{2\calI_{N}} l(l+1) \; .  
\eeq
Numerical values for $\calI$ and $\bra l^2\ket$ are shown 
in Table~\ref{tbl:Eint}, and the rotational energies $E_{N\,l}$ for the 
$N =2$ and $N = 1$ bands are shown in Table~\ref{tbl:Erot} .  
We emphasize that the energy subtraction due to angular momentum 
fluctuation, the second term of (\ref{Erot}), is very 
important to make the theoretical masses down close 
to experimental values.  
This is particularly important for the first $1/2^+$ excited state 
(the Roper).  

\begin{figure}[tbph]
   \vspace*{1cm}
   \centering
   \footnotesize
   \epsfxsize = 10cm
   \epsfbox{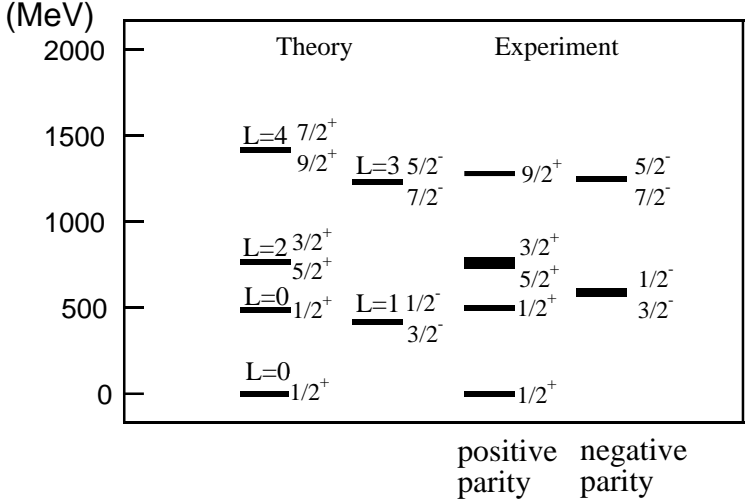} 
% \centerline{\protect
% \hbox{
% \psfig{file=f1BS.eps,height=3.cm,width=13.0cm,angle=-90}}}
 \begin{minipage}{12cm}
   \centering 
   \caption{ \small 
   Experimental nucleon mass spectra as compared with the DOQ 
   predictions.    \label{nspect}}
\end{minipage}
\end{figure}

Coupling the intrinsic spin $s$ of three quarks with the orbital
angular momentum $l$, take for example $s=1/2$, we consider
mass spectra of the $N=2$ and $N=1$ bands 
for nucleon excited sates as shown in Fig.~\ref{nspect}.
On the left hand side we show the theoretical results: one for the
positive parity states of $l=0, 2, 4, \cdots$ and the other
for the negative parity states of $l=1, 3, \cdots$.
In the theoretical side, two spin states $j = l \pm 1/2$
degenerate when
spin-orbit coupling is ignored, as experimental data suggest.
On the right hand side, experimental masses of well observed nucleon
excited states with four stars are shown~\cite{PDG1}.
One exception is the $5/2^{-}$ state of $D_{15}(2200)$ with two 
stars.  
This state is very likely to form a spin doublet with $G_{17}(2190)$.  
We do not list all the states but those which are well
identified with the $^2 8$ representations ($s = 1/2$)
of the spin-flavor group, and are well compared with the
DOQ model predictions.
Our theoretical formula for a fixed $N$ ($N=2$ for positive and $N=1$ 
for negative parity states) corresponds to the rigid rotor 
approximation.
Theoretically, one may consider effects from higher rotational bands 
of $N \ge 3$ in order to account for the softness of the intrinsic 
state.   
This is discussed in Ref.~\cite{tatoho2}.  

%%   Fig. ?   %%%%%%%%%%%%%%%%%%%%%%%%%%%%%%%%%%%%%%%%%%%%%%%%%%%%%
%%   Fig. ?   %%%%%%%%%%%%%%%%%%%%%%%%%%%%%%%%%%%%%%%%%%%%%%%%%%%%%

The mass formula (\ref{Erot}) should be compared with excitation 
energies of baryons. 
Hence it contains essentially one parameter, which is the 
oscillator parameter $\omega$.
It is determined here by the average mass splitting
between the first excited states of $1/2^+$ (Roper like states)
and the corresponding ground states
for the flavor SU(3) baryons (for details of SU(3) baryons, see the
discussion below).
The resulting value is $\omega = 644$ MeV~\cite{tatoho}.
Considering the simplicity of the DOQ model,
it is remarkable that many observed states fit very well
to the theoretical rotational band.
In particular we note that the first $1/2^{+}$ excited state of 
the Roper resonance emerges as the band head of the 
$N=2$ rotational band.  

If the fundamental structure of the excitation spectrum is produced by
gluon dynamics of quark confinement, 
we should see a similar pattern in other members of the spin and flavor 
multiplets.  
In a recent publication, we have analyzed in detail the mass spectrum 
of SU(3) baryons and found that 
the picture of deformed baryons works 
extremely well for a wide class of baryons~\cite{hotato,tatoho}.  
We have been able to explain almost 80 \% of the observed baryons as 
they fit into the rotational bands of the DOQ 
model.  

%------------------------------------------------------------------
\subsection{Three quark states}
%------------------------------------------------------------------
Baryon wave functions consist of spatial, spin, 
isospin and color parts.  
Here 
we discuss mainly spatial wave functions with definite permutation 
symmetry and with a good angular momentum.  
Since detailed method is found in Ref.~\cite{bhaduri}, 
we summarize 
minimally what we need in the computation of transition 
amplitudes.  

First we consider single particle wave functions 
in the deformed oscillator potential.  
They are given as products of one-dimensional harmonic oscillator wave 
functions $\psi_n(\vec x)$:
\beq
\label{sigle_psi}
\psi_{n_x n_y n_z}(\vec x) 
= \psi_{n_x}(x) \psi_{n_y}(y) \psi_{n_z}(z) \; .
\eeq
Since we are interested in excited states of prolate shape, 
we set $n_x = n_y = 0$, and write (\ref{sigle_psi}) as 
$\psi_{00\nu}(\vec x) = \psi_{0}(x) \psi_{0}(y) \psi_{\nu}(z)$.  

The state with a definite permutation symmetry for a three quark 
system, 
after removing the center of mass coordinates,  
is given by the product of the single particle wave functions 
$\psi_{00n}(\vec \rho)$ and 
$\psi_{00n^\pr}(\vec \lambda)$ with the constraint $n + n' = N$.  
They are either totally symmetric ($S$) or 
mixed symmetric ($MS$)
%% revised by takayama
\footnote{For the prolate deformation, there is no 
totally antisymmetric wave function 
like 
\( \Psi^{N A}\!\!\!\!=\!\!\!\!\phi _{100}(\rho)\phi _{001}(\lambda) 
\!\!-\!\! \phi 
    _{001} (\rho)\phi _{100}(\lambda) \), 
since quarks excite along only one direction, where 
($n_{x} = n_{y} = 0, n_{z} \neq 0)$.} 
.
The $MS$ states are further classified into 
two types of $\rho$ and $\lambda$ symmetries.  
Here we summarize single particle wave functions 
and the resulting intrinsic states for deformed baryons, 
$\Psi^{N\sigma}(\vec \rho, \vec \lambda)$, where 
$\sigma$ represents a permutation symmetry ($S$, $\lambda$ or $\rho$).  
In the following we introduce the parameter 
$\alpha = \sqrt{m \omega}$ and the deformation parameter \( d = 
\omega_{x}/\omega_{z} \).  
As we see in the previous section, \( d \) becomes 
\( N+1 \) after the energy minimization.  However, we keep it 
in the following expressions of wave functions to compute transition 
amplitudes as functions of the deformation parameter \( d \).  

\begin{itemize}
    \item Single particle wave function
    \begin{equation}
        \begin{split}
            \psi_{000}(\vec x) &= 
            \left(\frac{\alpha^{2}}{\pi} \right)^{\frac{3}{4}}
            \exp( - {\textstyle \frac{\alpha^2}{2}} 
            \{d^{\frac{1}{3}}(x^2+y^2)+d^{-\frac{2}{3}}z^2\} )   \\
            \psi_{001}(\vec x) &= 
            \left(\frac{\alpha^{2}}{\pi} \right)^{\frac{3}{4}}
            \sqrt{2}d^{-\frac{1}{3}}\,\alpha \,z
            \exp( - {\textstyle \frac{\alpha^2}{2}} 
            \{d^{\frac{1}{3}}(x^2+y^2)+d^{-\frac{2}{3}}z^2\} )   \\
            \psi_{002}(\vec x) &= 
            \left(\frac{\alpha^{2}}{\pi} \right)^{\frac{3}{4}}
            \sqrt{2} \bigl(
            d^{-\frac{2}{3}}\alpha^{2} z^{2} - \frac{1}{2}
            \bigr)
            \exp( - {\textstyle \frac{\alpha^2}{2}} 
            \{d^{\frac{1}{3}}(x^2+y^2)+d^{-\frac{2}{3}}z^2\} )   
        \end{split}
        \label{spwf}
    \end{equation}
    \item Three quark state
    \begin{itemize}
        \item \( N=0 \) (Spherical ground state) : \(d=1 \)
        \begin{equation}
            \Psi^{0S}(\vec \rho, \vec \lambda) 
            = \psi_{000}(\vec \rho) \psi_{000}(\vec \lambda)
            \label{Psi0}
        \end{equation}
        \item \( N=1 \) (Negative parity excitations) : \(d=2 \)
        \begin{equation}
            \begin{split}
                \Psi^{1\rho}(\vec \rho, \vec \lambda) 
                &= \psi_{001}(\vec \rho) \psi_{000}(\vec \lambda) 
                \\
                \Psi^{1\lambda}(\vec \rho, \vec \lambda) 
                &= \psi_{000}(\vec \rho) \psi_{001}(\vec \lambda)
            \end{split} 
            \label{Psi1}
        \end{equation}
        \item \( N=2 \) (Positive parity excitations) : \(d=3 \)
        \begin{equation}
            \begin{split}   
                \Psi^{2S}(\vec \rho, \vec \lambda) 
                &= \frac{1}{\sqrt{2}}
                \left( \psi_{002}(\vec \rho) \psi_{000}(\vec \lambda) 
                +  \psi_{000}(\vec \rho) \psi_{002}(\vec \lambda) \right) 
                \\
                \Psi^{2\rho}(\vec \rho, \vec \lambda)
                &= \psi_{001}(\vec \rho) \psi_{001}(\vec \lambda)  
                \\
                \Psi^{2\lambda}(\vec \rho, \vec \lambda) 
                &= \frac{1}{\sqrt{2}} 
                \left( \psi_{002}(\vec \rho) \psi_{000}(\vec \lambda) 
                -  \psi_{000}(\vec \rho) \psi_{002}(\vec \lambda) \right) 
            \end{split}
            \label{Psi2}
        \end{equation}  
    \end{itemize}
\end{itemize}

Since 
these deformed intrinsic states (\ref{Psi1}) and (\ref{Psi2}) 
are not eigenstates of angular 
momentum, we need to project out the states 
with good angular momentum for physical baryons.  
The projection method is conveniently performed first by expanding the 
deformed single particle state $\psi_{00\nu}(\vec x)$ 
by the wave functions $\phi_{nl,m=0}$ of the spherical three 
dimensional harmonic oscillator: 
\beq
\label{expandA}
\psi_{00\nu}(\vec x) = \sum_{nl} C_{nl}^{\nu} \phi_{nl,m=0}
\equiv \sum_{nl} C_{nl}^{\nu} \phi_{nl}  \; .
\eeq
The constraint $m=0$ reflects the axial symmetry around the $z$-axis.  
Substituting (\ref{expandA}) in Eqs. (\ref{Psi1}) and  (\ref{Psi2}), 
we find for three quark states $\Psi$:
\beq
\label{expandPsi}
\Psi^{N\sigma}(\vec \rho, \vec \lambda) 
= \sum_{n_\rho l_\rho  n_\lambda l_\lambda }
 F^{N\sigma}_{n_\rho l_\rho n_\lambda l_\lambda} \; 
      \phi_{n_\rho l_\rho }(\vec \rho) 
      \phi_{n_\lambda l_\lambda}(\vec \lambda) \; ,
\eeq
where the coefficients 
$F^{N\sigma}_{n_\rho l_\rho n_\lambda l_\lambda}$ are given by
\begin{equation}
\begin{split}
    % \label{F1rho}   
    F^{1\rho}_{n_\rho l_\rho n_\lambda l_\lambda} & =
    C^1_{n_\rho l_\rho} C^0_{n_\lambda l_\lambda} \; , \\ 
    % \label{F1lambda}
    F^{1\lambda}_{n_\rho l_\rho n_\lambda l_\lambda}
    &=
    C^0_{n_\rho l_\rho} C^1_{n_\lambda l_\lambda} \; , \\
    % \label{F2symm}
    F^{2S}_{n_\rho l_\rho n_\lambda l_\lambda}
    &=
    \frac{1}{\sqrt{2}} 
    \left( C^2_{n_\rho l_\rho} C^0_{n_\lambda l_\lambda}
    + C^0_{n_\rho l_\rho} C^2_{n_\lambda l_\lambda} \right) \; , \\
    %\label{F2rho}
    F^{2\rho}_{n_\rho l_\rho n_\lambda l_\lambda}
    &=
    C^1_{n_\rho l_\rho} C^1_{n_\lambda l_\lambda} \; , \\
    %\label{F2lambda}
    F^{2\lambda}_{n_\rho l_\rho n_\lambda l_\lambda}
    &=
    \frac{1}{\sqrt{2}} 
    \left( C^2_{n_\rho l_\rho} C^0_{n_\lambda l_\lambda}
    - C^0_{n_\rho l_\rho} C^2_{n_\lambda l_\lambda} \right) \; .
\end{split}
\end{equation}

The states with definite angular momentum $l,m$ can readily 
be projected 
out by operating the projection operator~\cite{PeiYocc57,RinShu80},
\beq
\label{defP}
\hat P_{lm} = \int d[A] D_{m0}^{l*}(A) 
\calR_{\rho}(A) \calR_{\lambda}(A) \; , 
\eeq
where $\calR_{\rho}(A)$ and $\calR_{\lambda}(A) $ are the rotation 
operator acting on the $\rho$ and $\lambda$ variables, and $A$ 
denotes a set of Euler angles for rotation.  
Applying $\hat P_{lm}$ to the deformed intrinsic state 
(\ref{expandPsi}), we find
\beq
\label{proj_gen}
\Psi^{N\sigma}_{lm} \equiv
\hat P_{lm} \Psi^{N\sigma}
=
\calN \sum_{n_\rho l_\rho n_\lambda l_\lambda}
F^{N\sigma}_{n_\rho l_\rho n_\lambda l_\lambda} \; 
%takayama..
%       (l_{\rho} 0 l_\lambda 0 | l0) \;         
\CG{l_{\rho}}{0}{l_\lambda}{0}{l}{0}
      [\phi_{n_\rho l_\rho } 
      \times \phi_{n_\lambda l_\lambda}]^{lm} \; ,
\eeq
where 
$[\phi_{n_\rho l_\rho } \times \phi_{n_\lambda l_\lambda}]^{l m}$
expresses that 
$\phi_{n_\rho l_\rho }$ and 
$\phi_{n_\lambda l_\lambda}$
are coupled to the state with total angular momentum $(l,m)$, 
and 
\( \CG{l_{\rho}}{0}{l_\lambda}{0}{l}{0} \)
are the standard Clebsh-Gordan coefficients.  
The normalization constants $\calN$ 
in (\ref{proj_gen}) are given by 
\beq
\calN^{-2} = \sum_{n_\rho l_\rho n_\lambda l_\lambda}
\vert F^{N\sigma}_{n_\rho l_\rho n_\lambda l_\lambda} 
\CG{l_{\rho}}{0}{l_\lambda}{0}{l}{0}
\vert^{2} \, . 
\eeq
Here we have performed angular momentum projection after taking
variation  (PAV) with respect to \( \omega \)'s.  
One could get better wave function
through angular momentum projection before variation (PBV). 
We have checked the PBV and found that the difference between 
the two projection schemes is not very significant up to \( l=2 \).

Finally to close this section, 
we write spatial-spin-flavor wave functions 
which are totally symmetric.  
Let us denote spin and isospin wave functions by $\chi$ and $\phi$, 
respectively, with a superscript
$\rho$, $\lambda$ or $S$.  
For nucleon states (denoted by $N$) with isospin 1/2, 
there are three possible states 
for a given spin and parity $j^P (P = (-1)^l)$:
\begin{equation}
    \begin{split}   
        {(\rm N1)} &\; \; \; 
        |N; [l_{S},\, 1/2]^j \ket\; =\;  \frac{1}{\sqrt{2}}
        \left( 
        [\Psi_l^{NS}, \, \chi^\rho ]^j \phi^\rho
        + [\Psi_l^{NS}, \, \chi^\lambda ]^j \phi^\lambda \right)\; ,     
        \\
        {\rm (N2)} &\; \; \;  
        |N; [l_{MS},\, 1/2]^j \ket \; =\;  \half
        \left( 
        [\Psi_l^{N\rho}, \, \chi^\rho ]^j \phi^\lambda
        + [\Psi_l^{N\rho}, \, \chi^\lambda ]^j \phi^\rho \right. 
        \\
        &  \hspace{3cm}
        \left.
        + \; \;  [\Psi_l^{N\lambda}, \, \chi^\rho ]^j \phi^\rho 
        - [\Psi_l^{N\lambda}, \, \chi^\lambda ]^j \phi^\lambda \right) \; ,
        % \label{Ntwo}
        \\
        {\rm (N3)} &\; \; \; 
        |N; [l_{MS},\, 3/2]^j \ket \;  =\; \frac{1}{\sqrt{2}}
        \left( 
        [\Psi_l^{N\rho}, \, \chi^S ]^j \phi^\rho
        + [\Psi_l^{N\lambda}, \, \chi^S ]^j \phi^\lambda \right)\; .
%       \label{Nthree} 
    \end{split}
    \label{None}
\end{equation}

On the left hand side, the notation is such that 
the orbital wave function is labeled simply by the angular 
momentum \( l \) and permutation symmetry \( \sigma \), $l_\sigma$.
The orbital angular momentum is then coupled with the 
three quark spin $s$ = 1/2 or 3/2 
to yield the total nucleon spin $j$.  
For deltas with isospin 3/2, there are two possible states for a 
given $j^P$:
\begin{equation}
    \begin{split}
        {(\Delta 1)} &\; \; \; 
        |\Delta; [l_{S},\, 3/2]^j \ket  \; =\;  
        [\Psi_l^{NS}, \, \chi^S ]^j \phi^S
%       \label{Done} 
        \, , \\
        {(\Delta 2)} &\; \; \; 
        |\Delta; [l_{MS},\, 1/2]^j \ket  \; =\;  \frac{1}{\sqrt{2}}
        \left( 
        [\Psi_l^{N\rho}, \, \chi^\rho ]^j 
        + [\Psi_l^{N\lambda}, \, \chi^\lambda ]^j \right)\phi^S\; .
    \end{split}
    \label{Dtwo}
\end{equation}

%==================================================================
%SECTION{Electromagnetic Transitions}
\setcounter{equation}{0}
\renewcommand{\theequation}{\arabic{section}-\arabic{equation}}
\section{Electromagnetic Transitions}
%==================================================================

%------------------------------------------------------------------
\subsection{Helicity amplitudes for the pion photoproduction}
%------------------------------------------------------------------

\begin{figure}[tbp]
   \vspace*{1cm}
   \centering
   \footnotesize
   \epsfxsize = 10cm
   \epsfbox{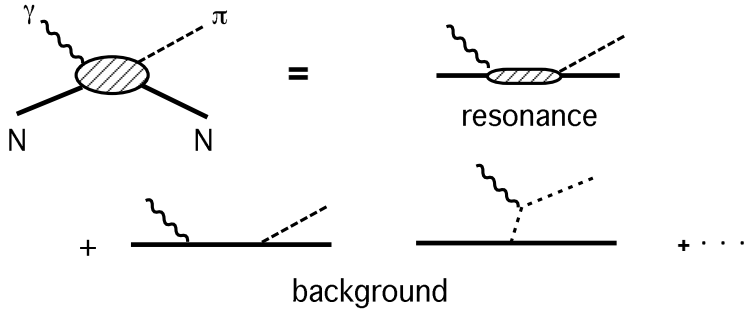} 
% \centerline{\protect
% \hbox{
% \psfig{file=f1BS.eps,height=3.cm,width=13.0cm,angle=-90}}}
 \begin{minipage}{12cm}
   \centering 
   \caption{ \small 
   Pion photoproduction which is decomposed into resonance and
background 
   terms.   \label{piprod}
}
\end{minipage}
 \end{figure}

Experimentally, electromagnetic 
couplings of excited baryon are extracted from single 
pion photoproductions (Fig.~\ref{piprod}).  
In a resonance dominant model, 
production amplitudes are assumed to be decomposed into the Born terms 
and resonance contributions.  
A distinguished feature of this process
is that not only the magnitude but also 
the sign of the resonance contributions can be determined relative 
to the sign of the Born terms~\cite{walker,moorhouse,PDG3}.  
More precisely, the sign $\epsilon$ in the following 
combination can be determined experimentally:
\beq
\bra N |H_\pi |N^*\ket \cdot \bra N^* | H_\gamma | N\ket \, 
\equiv 
\epsilon \vert 
\bra N |H_\pi |N^*\ket \cdot \bra N^* | H_\gamma | N\ket \vert  \; , 
\eeq
where $H_\pi$ and $H_\gamma$ are the interaction Hamiltonian for the 
pion and photon couplings.  
The sign $\epsilon$ is then included in the 
electromagnetic couplings $\bra N^* | H_\gamma | N\ket$.  
Therefore, in theoretical calculations, both the electromagnetic and 
pion couplings have to be calculated simultaneously.  
Ignorance of the pion coupling part might lead to incorrect 
results for the sign.  

%%   Fig. ?   %%%%%%%%%%%%%%%%%%%%%%%%%%%%%%%%%%%%%%%%%%%%%%%%%%%%%
%%   Fig. ?   %%%%%%%%%%%%%%%%%%%%%%%%%%%%%%%%%%%%%%%%%%%%%%%%%%%%%

Usually experimental amplitudes ($T$-matrix) 
are presented in the helicity basis, 
which are given as matrices in the space 
spanned by the initial and final helicity 
states~\cite{walker,moorhouse}.  
They are also functions of the scattering angles $\theta$ and $\phi$.  
The helicity amplitudes are then expanded by multipoles 
as~\cite{JacWick}
\beq
\label{Aexpand}
A_{\mu \lambda}(\theta, \phi)
=
\sum_{j} (2j+1) \, A^j_{\mu \lambda}\, 
d^j_{\lambda \mu} (\theta) \, e^{i(\lambda - \mu)\phi} \; .
\eeq
Here $\lambda = \lambda_\gamma - \lambda_i$ and 
$\mu = \lambda_\pi - \lambda_f = - \lambda_f$
are the initial and final state helicities with 
$\lambda_\gamma$, $\lambda_i$, $\lambda_\pi$ and 
$\lambda_f$ being the helicity of the photon, of the initial nucleon, 
of the pion and of the final nucleon, respectively.   
In the pion photoproduction,  
$\lambda$ takes four values of $-3/2, -1/2, +1/2$  
and +3/2, while $\mu = -1/2$ and +1/2.  
Among the eight components of $A_{\mu \lambda}$,  four of them are 
independent due to time reversal symmetry.  
The expansion coefficients $A^j_{\mu \lambda}$ which were
called the helicity elements in the literature~\cite{walker} 
contain  dynamical information of the excited 
states of the total spin $j$.  
The helicity elements $A^j_{\mu \lambda}$ in its form, however,  
do not have a definite parity.     
Parity eigenstates are projected out by the following 
linear combinations~\cite{walker,moorhouse,JacWick}:
\beq
\label{Clmd1}
C^{l_{\pi}+}_{\lambda} &=& \frac{1}{\sqrt{2}} 
  \left( A^j_{1/2\;  \lambda} + A^j_{-1/2\;  \lambda} \right)   \; , \\
\label{Clmd2}
C^{(l_{\pi}+1)-}_{\lambda} &=& \frac{1}{\sqrt{2}} 
  \left( A^j_{1/2\;  \lambda} - A^j_{-1/2\;  \lambda} \right) \; ,  
\eeq
where the total spin $j$ is given by $j = l_{\pi}+1/2$, and the 
parity is $P = -(-)^{l_{\pi}}$ for $C^{l_{\pi}+}_{\lambda}$ and 
$P = (-)^{l_{\pi}}$ for 
$C^{(l_{\pi}+1)-}_{\lambda}$.  
Here $l_{\pi}$ is the orbital angular momentum of the pion in the 
final state.  

Resonance parameters are determined by fitting the experimental 
amplitudes (\ref{Clmd1}) and (\ref{Clmd2}) 
by a simple Breit-Wigner form assuming resonance dominance~\cite{salin}
\footnote{Note that the expression the resonance propagator 
\eqref{Cres} differs from that given in Refs.~\cite{moorhouse, PDG3}.
However, \eqref{Cres} agrees with, for instance, eq. (2-18) of 
Ref.~\cite{moorhouse}.}:
\beq
C^{l_{\pi}\pm}_{\lambda}(W) &=& - \sum_{N^*} \epsilon
\left( \frac{\Gamma^\lambda_\gamma \Gamma_\pi}{kq} \right)^{1/2}
\; 
\frac{M}{W^2 - M^{*2} - i M \Gamma}
+ {\rm background} \nonumber \\
&\to& i \; \epsilon 
\left( \frac{\Gamma^\lambda_\gamma \Gamma_\pi}{kq \Gamma^2} 
\right)^{1/2} \; , 
\label{Cres}
\eeq
where $\Gamma$ is the total decay width of the resonance 
%takayama..
% $R$,
\(  N^*\) ,
$\Gamma^{\lambda}_{\gamma}$ and 
$\Gamma_{\pi}$ are the partial width for gamma (with the helicity 
$\lambda$) and pion emission.  
$k$ and $q$ are the momentum carried by the photon and the pion, 
respectively.  
In the second line of (\ref{Cres}) 
the CM energy $W$ is set to the resonance 
energy $M^{*}$, where the single resonance 
% $R$
\( N^{*} \)
 is assumed to dominate
the 
amplitude.  
The relative sign of each term is included in the parameter 
$\epsilon = \pm 1$.  
The helicity amplitudes for the electromagnetic couplings 
$A_\lambda^{j^{P}}$ are then defined by the imaginary part of the 
amplitude:
\beq
\label{Almd}
A_\lambda^{j^{P}} = 
-i \left( \frac{1}{(2j+1)\pi} \frac{k}{q} \frac{M}{M^*} 
\frac{\Gamma_\pi}{\Gamma^2} \right)^{-1/2}\, 
C^{l_{\pi}\pm}_{\lambda}(M^{*}) 
\; C_{\pi N}\, , \; \; \; \; 
j = l_{\pi}\pm 1/2\, , \; P=-(-)^{l_{\pi}} \, , 
\eeq
where $C_{\pi N}$ is the isospin Clebsh-Gordan coefficient for the decay of 
$N^*$ into the relevant $\pi N$ charge state~\cite{PDG2}.  
In this equation, the strength of the pion coupling in
$C^{l\pm}_{\lambda}$ 
is removed by multiplying the factor $\Gamma_{\pi}^{-1/2}$.   

Theoretically, the amplitude $C^{l_{\pi}\pm}_{\lambda}(M^{*})$ may be 
computed as
\beq
C(W) &=& i\; \bra N | H_\pi | N^* \ket 
\frac{M}{W^2 - M^{*2} - i M \Gamma} \bra N^* | H_\gamma |N \ket
\nonumber \\
&\xrightarrow{W \to M^{*}} & 
 \bra N | H_\pi | N^* \ket \, \bra N^* | H_\gamma |N \ket \; 
\frac{1}{\Gamma} \, .
\eeq%
{}From this and (\ref{Almd}), one obtains an alternative expression 
\beq
\label{defAh}
A_\lambda =   -\epsilon\, (KF)\; (-i)^l \; 
 \bra N^* | H_\gamma | N\ket \, , 
\; \; \; 
(KF) = \sqrt{ \frac{E_{N}(k)}{2\omega M} } \, , 
\eeq
where $l$ is an orbital angular momentum of an excited quark in $N^*$,
 and 
$\epsilon$ is the sign extracted from, together with another phase 
factor $(-i)^l$,  
the pion matrix element as given in Appendix B.   
At the end of this subsection, 
we summarize selection rules for various amplitudes and 
excited states in Table~\ref{varamp}.  

\begin{table}[t]
        \centering
        \footnotesize
        \caption{Relation between various amplitudes and quark model states.
        \label{varamp}
        }
        \vspace*{0.5cm}
        \begin{tabular}{c c|c c| c c c }
                \hline
                \multicolumn{2}{c |}{Excited states} & 
                \multicolumn{2}{c |}{Photon couplings} &
                \multicolumn{3}{c }{Pion photoproduction} \\
                \hline
                $j^P$ & $[l,\, s]^j$ & Multipole & Helicity 
                     & $l_{\pi}$ & Multipole & Helicity  \\
                \hline
                $1/2^+$ & $[0,\, 1/2]^{1/2}$ & $M1$ & $A_{1/2}$ 
                     & 1 & $M_{1^{-}}$ &  $C^{1^-}_{1/2}$  \\
                        & $[2,\, 3/2]^{1/2}$ &   &   &   &   &  \\
                $3/2^+$ & $[2,\, 1/2]^{3/2}$ & $M1, E2$ & $A_{1/2,3/2}$ 
                     & 1 & $M_{1^{+}},E_{1^{+}}$ &  $C^{1^+}_{1/2,3/2}$  \\
                        & $[0,\, 3/2]^{3/2}$ & $M1$ &  &   &   &  \\
                $5/2^+$ & $[2,\, 1/2]^{5/2}$ & $M3, E2$ & $A_{1/2,3/2}$ 
                     & 3 & $M_{3^{-}},E_{3^{-}}$ &  $C^{3^-}_{1/2,3/2}$  \\
                        & $[4,\, 3/2]^{3/2}$ & $M3$ &  &   &   &  \\
        \hline
                $1/2^-$ & $[1,\, 1/2]^{1/2}$ & $E1$ & $A_{1/2}$ 
                     & 0 & $E_{0^{+}}$ &  $C^{0^+}_{1/2}$  \\
                        & $[1,\, 3/2]^{1/2}$ &    &   &   &   &  \\
                $3/2^-$ & $[1,\, 1/2]^{3/2}$ & $M2, E1$ & $A_{1/2,3/2}$ 
                     & 2 & $M_{2^{-}},E_{2^{-}}$ &  $C^{2^-}_{1/2,3/2}$  \\
                        & $[3,\, 3/2]^{3/2}$ & $M2$ &  &   &   &  \\
                $5/2^-$ & $[3,\, 1/2]^{5/2}$ & $M2, E3$ & $A_{1/2,3/2}$ 
                     & 2 & $M_{2^{+}},E_{2^{+}}$ &  $C^{2^+}_{1/2,3/2}$  \\
                        & $[1,\, 3/2]^{3/2}$ & $M2$ &  &   &   &  \\
                \hline
        \end{tabular}
\end{table}

%------------------------------------------------------------------
\subsection{Photon couplings}
%------------------------------------------------------------------

\begin{figure}[tbp]
   \vspace*{1cm}
   \centering
   \footnotesize
   \epsfxsize = 7cm
   \epsfbox{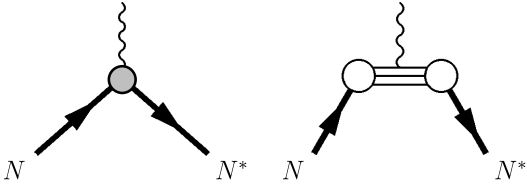} 
% \centerline{\protect
% \hbox{
% \psfig{file=f1BS.eps,height=3.cm,width=13.0cm,angle=-90}}}
 \begin{minipage}{10cm}
   \centering 
   \caption{ \small 
   A photon coupling to a quark in baryons.   \label{qtrans}
}
\end{minipage}
 \end{figure}

The electromagnetic coupling is given 
by the interaction hamiltonian 
\beq
\label{Hem}
H_{\gamma} = - e \int d^{3}x \, \vec J  \cdot \vec A\; ,
\eeq
where $\vec A$ is the photon field, and for the current $\vec J$ we 
adopt the non-relativistic form 
\beq
\vec J &=&\sum_{j=1}^{3} \vec J^{(j)} = 3 \vec J^{(3)} \, ,
\nonumber
\\
\label{Jem}
\vec J^{(3)} &=& \frac{1}{2m}
\left( u_f^\dagger ( i \lnabla - i \rnabla ) u_i
+
\rnabla \times (u_f^\dagger \vec \sigma u_i) 
\right) \tau_\mu \; .
\eeq
Here we have used overall permutation symmetry of quarks in the 
baryons, and
$u_i$ and $u_f$ are the 
two component Pauli spinors for the initial and final 
state quarks, respectively (see Fig.~\ref{qtrans}).  
The isospin matrix $\tau_\mu$ is either $\tau_0 = 1$ for 
the isoscalar current  and $\tau_i = \tau_3$ for the 
isovector current.  
We will consider relativistic corrections later.  

%%   Fig. 3   %%%%%%%%%%%%%%%%%%%%%%%%%%%%%%%%%%%%%%%%%%%%%%%%%%%%%
%%   Fig. 3   %%%%%%%%%%%%%%%%%%%%%%%%%%%%%%%%%%%%%%%%%%%%%%%%%%%%%

Helicity amplitudes $A_{1/2,3/2}$ are then computed as given in 
(\ref{defAh}):
\beq
\label{Aone}
A_{1/2} &=& -\epsilon (-i)^l \, (KF) \; 
\bra N^*(j,\, 1/2) | \vec J \cdot \vec A_{+1} |N(1/2,\, -1/2) \ket 
\; , \nonumber \\
\label{Athree}
A_{3/2} &=& -\epsilon (-i)^l\, (KF) \; 
\bra  N^*(j,\, 3/2) | \vec J \cdot \vec A_{+1} |N(1/2,\, 1/2)\ket 
\; .
\eeq
In order to relate the helicity amplitudes with the multipole
amplitudes, 
we consider the multipole expansion for the photon field~\footnote{
The same symbol $A$ is used for the photon field and helicity 
amplitudes, and $\epsilon$ for the photon 
polarization vector and the sign of the amplitudes.  
Both notations are 
standard and there will be no confusion in the following discussions.}
\beq
\label{helrep}
\vec A_\lambda = \vec \epsilon_\lambda e^{i\vec k \cdot \vec x} \; .
\eeq
Here $\vec \epsilon_\lambda$ 
is the polarization vector of helicity 
$\lambda = \pm 1$.  
The momentum $\vec k$ is related to the physical momentum $\vec
k_{phys}$ by 
\beq
\vec k = \sqrt{\frac{2}{3}} \vec k_{phys}  \; ,
\eeq
in the convention of coordinate transformation 
(\ref{cotrans}).  
If we choose the z-axis along the direction of photon propagation 
$\vec k$, the plane wave of a definite helicity 
(\ref{helrep}) can be expanded 
by multipoles as~\cite{rose}
\beq
\label{Aexpl}
\vec A_\lambda = -\sqrt{2\pi} \sum_{l=1}^\infty
i^l \sqrt{2l+1} 
\left(
\lambda \vec A^{(M)}_{l\lambda} + i 
\vec A^{(E)}_{l\lambda}   \right)  \; .  
\eeq
% \beq
% \label{Aexpl}
% \vec A_\lambda = \sqrt{2\pi} \sum_{l=1}^\infty
% i^l \sqrt{2l+1} 
% \left(
% \vec A^{(M)}_{l\lambda} + i \lambda 
% \vec A^{(E)}_{l\lambda}   \right)  \; .  
% \eeq

Here the magnetic ($M$) and electric ($E$) multipole fields are 
given by 
\beq
\vec A^{(M)}_{lm}(\vec r) &=& j_l(kr) \vec Y^l_{lm}(\hat r) \; , 
\; \; \; \; \; \; 
\vec Y^l_{lm} \; = \; \frac{1}{\sqrt{l(l+1)}} \vec L Y_{lm} (\hat r)
\; , 
\nonumber \\
\label{Afields}
\vec A^{(E)}_{lm}(\vec r) &=& 
- \frac{i}{k} \rnabla \times \vec A^{(M)}_{lm}(\vec r) \; , 
\eeq
where we have adopted the standard convention for the spherical
harmonics~\cite{VMK88},  and 
$j_l(kr)$ are the spherical Bessel functions.  

Substituting (\ref{Aexpl}) in (\ref{Hem}) and applying the 
Wigner-Eckart theorem to each multipole amplitude, we find the 
following formulae relating the helicity and multipole amplitudes:\\   
For  $j=l+3/2$:
\beq
\label{A1_1}
A_{1/2} &=& 
- \epsilon\, (KF)\,  \sqrt{2\pi} 
\sqrt{\frac{l+1}{2l+4}} \; \calM(M\, l+1) \nonumber \\
\label{A3_1}
A_{3/2} &=&  
-\epsilon\, (KF)\, \sqrt{2\pi}
\sqrt{\frac{l+3}{2l+4}} \; \calM(M\, l+1)
\eeq
For  $j=l+1/2$:
\beq
\label{A1_2}
A_{1/2} &=& 
\epsilon\, (KF)\, \sqrt{2\pi} \left(
+ \sqrt{\frac{l+2}{2l+2}} \; \calM(M\, l+1)
- \sqrt{\frac{l}{2l+2}} \; \calM(E\, l)  \right) \nonumber \\
\label{A3_2}
A_{3/2} &=& 
\epsilon\, (KF)\, \sqrt{2\pi} \left(
- \sqrt{\frac{l}{2l+2}} \; \calM(M\, l+1)
- \sqrt{\frac{l+2}{2l+2}} \; \calM(E\, l)  \right)
\eeq
For  $j=l-1/2$:
\beq
\label{A1_3}
A_{1/2} &=& 
\epsilon\, (KF)\, \sqrt{2\pi} \left(
+ \sqrt{\frac{l-1}{2l}} \; \calM(M\, l-1) 
+ \sqrt{\frac{l+1}{2l}} \; \calM(E\, l) \right) \nonumber \\
\label{A3_3}
A_{3/2} &=& 
\epsilon\, (KF)\, \sqrt{2\pi} \left(
+ \sqrt{\frac{l+1}{2l}} \; \calM(M\, l-1) 
- \sqrt{\frac{l-1}{2l}} \; \calM(E\, l) \right)
\eeq
For  $j=l-3/2$:
\beq
\label{A1_4}
A_{1/2} &=& 
- \epsilon\, (KF)\, \sqrt{2\pi} 
\sqrt{\frac{l}{2l-2}} \; \calM(M\, l-1) \nonumber \\
\label{A3_4}
A_{3/2} &=& 
+ \epsilon\, (KF)\, \sqrt{2\pi} 
\sqrt{\frac{l-2}{2l-2}} \; \calM(M\, l-1) 
\eeq
Writing $|N\ket \to |i\ket$ and 
$|N^*\ket \to |f\ket$,  the reduced matrix elements are defined by 
\begin{equation}
\begin{split}
\calM(M\,l) 
&\equiv i\, \bra f || \vec J \cdot \vec A_l^{\; (M)} || i \ket 
\cdot \bra f|\tau_{\mu}|i\ket \; , \\
\calM(E\,l) 
&\equiv i\, \bra f || \vec J \cdot \vec A_l^{\; (E)} || i \ket 
\cdot \bra f|\tau_{\mu}|i\ket \; , 
\end{split}
\label{defcalM}
\end{equation}
where the isospin part is the ordinary matrix elements.  
In these equations,  the phase $i$ is introduced to make the matrix
elements 
$\calM$ pure real.  
Furthermore, 
in Eqs. (\ref{A1_1}) -- (\ref{A3_4}), the phase factor $(-i)^l$ in 
(\ref{Aone}) is canceled by another factor $i^l$
coming from the multipole expansion (\ref{Aexpl}).

Let us briefly outline how to compute the multipole amplitudes 
(\ref{defcalM}).  
For illustration, we consider magnetic transitions $M\,l$.  
Let us write the reduced matrix element as 
\beq
\label{Mlone}
\bra f \;||\; \vec J \cdot \vec A_{l}^{(M)} \;||\; i \ket 
\to  \frac{1}{2m}
\left( i \; u_f^\dagger ( \lnabla - \rnabla ) u_i
+
\rnabla \times (u_f^\dagger \vec \sigma u_i) 
\right) \cdot \vec A_{l}^{(M)} \; ,
\eeq
where on the right hand side it is understood that 
the integral and sum are taken over configuration and spin spaces.  
The initial and final states are written as 
$u_i$ and $u_f$ in order to indicate the operation of derivatives.  
The isospin part can be treated separately 
in a trivial manner and is not shown here.
Now in (\ref{Mlone}) integrating the left derivative ($\lnabla$) 
terms by parts and using the relation
$\vec \nabla \cdot \vec A = 0$ and $\hat x \cdot \vec A^{(M)}$ = 0 
(the radial vector $\hat x$ appears when the derivative hits the 
spherical initial state which is the ground state nucleon), 
one finds that the convection ($\lnabla - \rnabla$) term 
vanishes.  
Thus only the spin term survives:
\beq
\label{Mltwo}
\calM (M\,l)
= \frac{i}{2m}\; u_f^\dagger \vec \sigma u_i
\cdot 
\vec \nabla \times \vec A_{l}^{(M)} \; .
\eeq
Using the relation between the magnetic and electric fields 
(\ref{Afields}), one arrives at the expression
\beq
\calM (M\,l) &=& \frac{i}{2m}
\label{Mlgeneral}
\left[
- \sqrt{\frac{l}{2l+1}} \; 
\bra f || [Y_{l+1}\sigma ]^l j_{l+1}|| i\ket  \right. 
\nonumber \\
& & \hspace{1cm}  \left. +\;\sqrt{\frac{l+1}{2l+1}} \; 
\bra f || [Y_{l-1}\sigma ]^l j_{l-1}|| i\ket 
 \right]
 \bra f | \tau_{\mu} |i \ket \; . 
\eeq

For electric  transitions $E\,l$, again integrating by parts, one finds 
\beq
\label{Elone}
\calM (E\,l)
=  \frac{1}{m} u_f^\dagger \,\rnabla\, u_i 
\cdot \vec A_{l}^{(E)} 
+ \frac{k}{2m} u_f^\dagger \,\vec \sigma \, u_i \; 
\cdot \vec A_{l}^{(M)} \; .
\eeq
Since the initial state is spherical, the first term can be written as 
\beq
(u_f^\dagger \,\rnabla\, u_i) \cdot \vec A_{l}^{(E)} 
= 
u_f^\dagger   u_i^\pr \; \hat x \cdot \vec A_{l}^{(E)} \; .
\eeq
Then using the following properties of the spherical harmonics, 
\beq
\hat x \cdot \vec A^{(E)}_{lm}
= \sqrt{l(l+1)}\;  \frac{j_l(kr)}{kr} \; Y_{lm} \; ,
\eeq
for the first term  and 
\beq
\vec \sigma \cdot \vec A^{(M)}_{lm} = 
\vec \sigma \cdot \vec Y^l_{lm}\;  j_l (kr)
= [Y_l, \sigma]^{lm}\;  j_l(kr)  \;,
\eeq
for the second term of (\ref{Elone}), one arrives at the expression 
\beq
\label{Elgeneral}
\calM(E\,l) =  -i
\left [
\frac{\sqrt{l(l+1)}}{m}
\bra f \; || \; Y_l \frac{j_l(kr)}{kr}|| i^\pr\; \ket 
+
\frac{k}{2m} \bra f \; || \; [Y_l \sigma]^l j_l (kr)|| \; i \ket
\right]
\, \bra f | \tau_{\mu} |i\ket \; , 
\eeq
where $|i^\pr\ket = \del / \del r |i\ket$.  
Equations (\ref{Mlgeneral}) and (\ref{Elgeneral}) are the general 
expressions for electromagnetic transitions $N \gamma \to N^*$.
Concrete expressions for various matrix elements are given in 
Appendix C.

%==================================================================
%SECTION{Results and discussions}
\setcounter{equation}{0}
\renewcommand{\theequation}{\arabic{section}-\arabic{equation}}
\section{Results and discussions}
%==================================================================

We have computed the matrix elements as given in  
(\ref{M_N1a}) -- (\ref{M_N3d}) for various excited baryon states 
of $N=1$ (negative parity) and $N=2$ (positive parity) bands\footnote{
We do not discuss the transition 
$\Delta(1232) \to N$, since in the DOQ model they both appear spherical 
states 
and the transitions are the same as for the spherical quark model.}. 
Matrix elements are then investigated as functions of the deformation 
parameter $d$.   
Note that the actual $d$'s, when determined by 
the energy minimization, are $d = 2$ for $N = 1$ and $d=3$ for $N = 2$ 
bands, respectively.   
We use the oscillator parameter as determined from the mass spectrum: 
$\omega = 644$ MeV.  
For the photon momentum $k$ we use the values computed from the 
theoretical masses of excited states.  
The use of the experimental values do not change the essential features 
in the following discussions.  

We have computed the amplitudes for all the DOQ model 
states which contain more states than experimentally observed.  
For a more realistic treatment, we should consider 
configuration mixings due to residual interactions.  
They could be induced by gluon and meson exchange interactions.  
For instance, a tensor force between quarks strongly mixes 
the two states of $^{2}P_{MS}$  and $^{4}P_{MS}$ for $1/2^-$ 
$N(1535)$ and $N(1650)$~\cite{isg,asy}.  

In Tables \ref{tbl:Aproton}-\ref{tbl:Adelta}, 
we summarize helicity amplitudes for photon couplings 
to nucleon and delta excitations.  
The first column is for the present results with 
suitable deformation ($d = 3$ for positive parity states and $d = 
2$ for negative parity states) and 
the second column is for the spherical limit ($d$ = 1) 
which corresponds to the conventional non-relativistic (NR) quark model.  
These theoretical values are then compared to experimental data 
listed in the third column, where we indicate observed states which 
are naively identified with one of the DOQ model states.   

%
%  Table A_{1/2,3/2} for Nucleon
%
%\newpage
\begin{table}[htbp]
        \centering
        \footnotesize
        \caption{Helicity amplitudes of nucleon excitations.
        \label{tbl:Aproton}
}
        \vspace{0.5cm}
        \begin{tabular}{ c c c|c c c|c c c | c}
\hline
\hline

 &  &  &  & $A_{1/2}$ &  &  & $A_{3/2}$&     \\
\multicolumn{3}{c|}{Positive parity}  
& $d=3$ & $d=1$ & Exp. & $d=3$ & $d=1$ & Exp. \\
\hline
\hline
1/2$^{+}$ & $^{2}S^{\prime}_{S}$ & $p$ &
 $+115$ & $+25$ & $-68\pm 5$ &  --  & --  & -- & $P_{11}(1440)$\\
&  & $n$ & $-74$ & $-16$ & $+39\pm 15$ &  --  & --  & -- & \\
& $^{2}S_{MS}$ & $p$ & $+16$ & $+17$ & $+5\pm 16$ & 
                    --  & --  & -- & $P_{11}(1710)$\\
& & $n$ & $-5$ & $-6$ & $-5\pm 23$ & --  & --  & --  & \\
& $^{4}D_{MS}$ & $p$ & 0     &  0    &  & --  & --  & --  & \\
&  & $n$ &  $+3$  & $+4$  & $-5\pm 23$  &  --  & --  & --  &  \\
\hline
3/2$^{+}$ & $^{2}D_{S}$& $p$ & 
  $+70$ & $+113$ & $+52\pm 39$ & $-22$ & $-36$ & $-35\pm24$ & $P_{13}(1720)$\\
 & & $n$ & $-21$ & $-34$ & $-2 \pm 26$ & 0 & 0 & $-43\pm 94$ &  \\
 & $^{2}D_{MS}$ & $p$ & $+57$ & $+80$ &  & $-18$ & $-25$ &  & \\
 &              & $n$ & $-40$ & $-56$ &  & $+18$ & $+25$ &  & \\
 & $^{4}D_{MS}$ & $p$ &  0    & 0     &  &    0  &  0    &  & \\
 &              & $n$ & $-6$ & $-8.5$ &  & $+11$ & $+15$ &  & \\
 & $^{4}S_{MS}$ & $p$ & 0     & 0     &  & 0     & 0     &  & \\
 &              & $n$ & $-12$ & $-14$ &  & $-20$ & $-23$ &  & \\
\hline
5/2$^{+}$ & $^{2}D_{S}$ & $p$ & 
 $+8$ & $+12$ & $-17\pm 10$ & $-44$ & $-71$ & $+127\pm 12$ &
$F_{15}(1680)$ \\
&  & $n$ & $-26$ & $-42$ & $+31\pm 13$ & 0 & 0 & $-30\pm 14$ & \\
& $^{2}D_{MS}$ & $p$ & $+6$  & $+9$ &  & $-36$ & $-50$ &  & \\
            &  & $n$ & $+15$ & $+21$ &  & $+36$ & $+50$ &  & \\
& $^{4}D_{MS}$ & $p$ & 0    & 0   &  & 0     & 0     &  & \\
            &  & $n$ & $-4$ & $-6$ &  & $-17$ & $-24$ &  & \\
\hline
\hline
 &  &  &  & $A_{1/2}$ &  &  & $A_{3/2}$&     \\
\multicolumn{3}{c|}{Negative parity} &  
$d=2$ & $d=1$ & Exp. & $d=2$ & $d=1$ & Exp. \\
\hline
\hline
1/2$^{-}$ & $^{2}P_{MS}$  & $p$ & $+142$   & $+154$   &  
        74$\pm$11 & -- & -- & -- & $S_{11}(1535)$\\
 &   & $n$ & $-117$  & $-126$  &  $-72\pm 25$ & -- & -- & -- & \\
& $^{4}P_{MS}$ & $p$  & 0     & 0     &  $48\pm 16$ & -- & -- & -- & 
            $S_{11}(1650)$\\
&  & $n$ & $+13$ & $+14$ & $-17\pm 37$ & -- & -- & -- & \\
\hline
3/2$^{-}$ & $^{2}P_{MS}$  & $p$ 
& $-18$ & $-19$ & $-23\pm 9$ & $-127$ & $-137$ & $163\pm 8$ & $D_{13}(1520)$\\
&  & $n$ & $+55$ & $+59$ & $-64\pm 8$ & $+127$ & $+137$ & $-141\pm 11$
& \\
& $^{4}P_{MS}$ & $p$ 
   & 0 & 0 & $-22\pm 13$ & 0 & 0 & $0\pm 19$  & $D_{13}(1700)$ \\
&  & $n$ & $-6$  & $-6$  & $0\pm 56$ & $-30$ & $-33$ &  $-2\pm 44$ &  \\
%             
%             
%             & $^{4}F_{MS}$  & 0     & -- &     & 0     & -- &   & \\
\hline
% 5/2$^{-}$ & $^{2}F_{S}$ & 73.9  & --  &  & -21.4  & -- &  & \\
%             & $^{2}F_{MS}$  & 0 & -- &  & 0 & -- & \\
5/2$^{-}$ & $^{4}P_{MS}$ & $p$ 
 & 0 & 0 & $+19\pm 12$ & 0 & 0  & $+19\pm 12$ & $D_{15}(1675)$ \\
& & $n$ & $-22$ & $-28$ & $-47\pm 23$ & $-31$ & $-39$ & $-69\pm 19$ & \\
% \hline
% 7/2$^{-}$ & $^{2}F_{S}$ & -22.1  & --  &  & 36.5  & -- &  &  \\
%             & $^{2}F_{MS}$  & 0 & -- &  & 0 & -- &   & \\
\hline
\hline

\end{tabular}
\end{table}

%
%  Table A_{1/2,3/2} for Deltas
%
%\newpage
\begin{table}[tp]
        \centering
        \caption{Helicity amplitudes of delta excitations
        \label{tbl:Adelta}
}
        \vspace{0.5cm}
\begin{tabular}{ c c|c c c|c c c | c }
\hline
\hline
 &  &  & $A^{1/2}$ &  &  & 
         $A^{3/2}$ &   \\
\multicolumn{2}{c|}{Positive parity}  & 
$d=3$ & $d=1$ & Exp. & $d=3$ & $d=1$ & Exp.   \\
\hline
\hline
1/2$^{+}$ & $^{2}S_{MS}$ & $+12$ & $+14$ &             & -- & -- & --
& \\
        & $^{4}D_{S}$   & $+11$ & $+18$ & $-12\pm 30$ & -- & -- & -- 
& 
        $P_{31}(1910)$\\
\hline
3/2$^{+}$ & $^{2}D_{MS}$ & $-9$  & $-14$ &   & $+11$  &$+18$&   &
\\
          & $^{4}D_{S}$  & $-12$ & $-24$ &   & $+20$  &$+41$&   &
 $P_{33}(1920)$\\
% sign change   $^{4}S^{\prime }_{S}$                    
          & $^{4}S^{\prime}_{S}$  
          & $-54$ & $-37$ & $-20\pm 29$ & $-94$ & $-65$ & $+1\pm 22$  &
                    $P_{33}(1600)$\\
\hline
5/2$^{+}$ & $^{2}D_{MS}$ & $+28$& $+46$ &   & $+22$ & $+36$  & &  \\
          & $^{4}D_{S}$ & $-8$ & $-16$ & $ 27\pm 13$  
                & $-33$ & $-65$ & $-47\pm 19$ & $F_{35}(1905)$ \\
\hline
\hline
 &  &  & $A^{1/2}$ &  &  & 
         $A^{3/2}$ &   \\
\multicolumn{2}{c|}{Negative parity}  &
 $d=2$ & $d=1$ & Exp. & $d=2$ & $d=1$ & Exp. \\
\hline
\hline
% sign change
1/2$^{-}$ & $^{2}P_{MS}$  & $-35$ & $-40$ & 19 $\pm$ 16
               & -- & -- & -- & $S_{31}(1620)$ \\
\hline
% sign change
3/2$^{-}$ & $^{2}P_{MS}$  
           & $+69$ & $78$ & $116\pm 17$ & $+69$ & $+78$ & 
                   $77 \pm 28$  & $D_{33}(1700)$ \\
\hline

\end{tabular}
\end{table}

As explained in the preceding subsection, we have computed both the 
photon and pion couplings simultaneously.  
Therefore, 
there is no ambiguity in relative signs of the photon couplings.  
In the nucleon channels $5/2^{+}(^{2}D_{S})$ and
$3/2^{-}(^{2}P_{MS})$, 
however, the signs of the 
present result do not agree with those of, for instance, the 
pioneering work by Feynman, Kislinger and Ravndal~\cite{FKR}.  
However, if we included the sign as tabulated by Moorhouse and 
Parsons~\cite{MoPa},  
and an additional phase factor $i^{l=2}$ for the $D$-wave component 
in the multipole expansion (\ref{Aexpl}), 
the present results agree with the results of FKR. 
The convention of the sign has not been considered seriously even in 
recent publications. In the work of Koniuk and Isgur~\cite{KI}, their 
convention was taken to reproduce empirical one, which has been 
followed by several literatures~\cite{bhaduri2,shimizu,meyer}.    
This situation was partly discussed by Capstick in Ref.~\cite{capstick}.

In many channels, the DOQ model ($d$ = 2 or 3)  
reproduces experimental amplitudes reasonably well including both 
signs and absolute values nearly to the same extent that the NR quark 
model ($d$ = 1) does.  
This means that the $d$ dependence is not very strong,  which 
is somewhat surprising, since excited states are rather 
strongly deformed as $d = 2$ and 3.  
A possible reason could be that the number of the 
relevant degrees of freedom for baryons, which are valence 
quarks, is not very large.    
If we, however, look closely at the numbers, 
the positive parity baryons have slightly stronger 
$d$ dependence than the negative parity baryons, as $d$ is larger for 
the former.  
Typically, as it is seen in the channel 
$3/2^{+}\; ^{2}D_{S}$, 
$A_{1/2}^{\rm proton} (d = 1) = 113 \times 10^{-3}\; 
\text{GeV}^{-1/2}$ and 
$A_{1/2}^{\rm proton} (d = 3) = 70 \times 10^{-3}\; 
\text{GeV}^{-1/2}$
as compared with the experimental value 
$A_{1/2}^{\rm proton} ({\rm exp}) = 52 \pm 39 \times 10^{-3}\; 
\text{GeV}^{-1/2}$.  

Theoretical predictions differ most significantly for the first 
$1/2^+$ excited state for the Roper resonance; 
not only the magnitude but also the sign do not agree with data.  
The relevant amplitude is $A_{1/2}$ or $M1$.  
Indeed the discrepancy in the sign is puzzling, if the $1/2^+$ 
excitation is simply considered as a radial excitation of the ground 
state nucleon.  
It does not matter whether it is a single particle or collective 
excitation.  
Due to an S-wave nature of the orbital motion, spin structure and 
hence the magnetic moment of the excited state should be 
the same as the ground state.  
Now we know that the magnetic moments of the nucleon, 
the delta~\cite{nefkens} and the M1 transition of the 
delta~\cite{satolee} are 
reasonably well explained by the quark model.  
Therefore, one would naively expect that the magnetic transition of
the 
$1/2^+$ excited state were explained as well as these transition
amplitudes in the above.  

The discrepancy in the sign is known for a long time and various 
ideas to overcome the problem have been proposed.  
In Ref.~\cite{bhaduri2}, deformation effects was considered to 
be one candidate of such.  
However, their conclusion was based on the wrong sign of the 
amplitude.  
In fact, the deformation is to increase the amplitude but with 
the wrong sign remained.    
Since the problem in the Roper resonance is in itself 
an interesting question, we discuss two important effects
in the next two subsections.  

In the end of this subsection, 
we comment on the sign of the amplitudes for the 
$N(1440)$  and $\Delta(1600)$.  
The latter is interpreted as the Roper resonance of the delta.  
Let us look at the formulae for the nucleon (\ref{M_N1a}) and 
the delta for $d = 1$ when
the sum in the radial integral reduces just to a single 
term.  
We have 
\beq
M1(N;  {}^2 S'_{S}) &=& 
\frac{3}{4\sqrt{2}} M1(\Delta ; {}^4  S'_{S}) \nonumber \\
&=& 
-\frac{k}{2m}\frac{1}{\sqrt{4\pi}} 
F^{2S}_{0010} (00|j_{0}|10) \cdot ({\rm isospin\; part})\, .  
\eeq

From this, after taking into account the sign of the pion 
matrix element (see Table~\ref{signpi} in Appendix B), 
we observe that the signs of the transitions for 
$N(1440)$ and $\Delta(1600)$  are different.  
This fact is correctly shown in Tables~\ref{tbl:Aproton} and 
\ref{tbl:Adelta}.  
In literatures, however, the signs of the two matrix elements are 
found to be the same~\cite{capstick,ClLi}.  

%------------------------------------------------------------------
\subsection{Diagonalization of the $1/2^+$ states}
%------------------------------------------------------------------

In the DOQ model, the two $1/2^+$ states for the nucleon and the 
Roper, $^{2}S^\prime_{S}(N=2)$ and $^{2}S_{S}(N=0)$  
are not orthogonal to each other, since they are eigenstates 
of different Hamiltonians with different deformation:   
$d$ = 1 and 3.  
Generally, those states with the same spin parity $j^P$ but with 
different $N$ are not orthogonal.  
Explicitly, denoting 
$|{}^{2}S^\prime_{S}(N=2)\ket , |{}^{2}S_{S}(N=0)\ket \equiv
|2\ket , |0\ket$, they 
have non-zero overlap 
\beq
\label{overlap02}
\bra 0 |2\ket \sim F^{2S}_{0000} \equiv \varepsilon \, .
\eeq
In other words, the two states $|0\ket$ and $|2\ket$ form a 
non-orthogonal basis.  
Hence we need to reconstruct orthonormal basis.  

Using the non-orthogonal states $|0\ket$ and $|2\ket$, the 
eigenvalue equation takes the following $2 \times 2$ matrix form:
\beq
& & (H - EN) \Psi = 0 \, , 
\eeq
where
\beq
& & \; \; \; 
H = \left( \begin{array}{c c}
           E_0 & \varepsilon E_{02} \\
           \varepsilon E_{02} &  E_2
           \end{array}
    \right) \, , \; \; \; \; 
N = \left( \begin{array}{c c}
           1 & \varepsilon \\
           \varepsilon  &  1
           \end{array}
    \right)\, ,  \; \; \; \; 
\Psi = \left( \begin{array}{c}
           a |0\ket \\
           b |2\ket
           \end{array}
      \right) 
    \equiv \left( \begin{array}{c}
           a  \\
           b 
           \end{array}
      \right) \, .
\eeq
Here the diagonal components of the energy matrix are
given by (\ref{Erot}):
\beq
E_N = \bra N | H | N \ket 
= E^{int}_{N} - \frac{1}{2\calI_{N}}\bra l^2  \ket_{N} \; ,
\eeq
while the off diagonal components are, for 
simplicity, taken as the left-right average:
$E_{02} =  \bra 0 | H | 2 \ket =
( E_0 + E_2)/2 \; .
$
Eigenvalues and the corresponding eigenstates are given by
\beq
E_{N,R} = E_{0} \mp (\gamma -1) \Delta  \;,
\eeq
and
\beq
\label{NRdiag}
|N \ket &= & \frac{-1}{\sqrt{2(1-\sqrt{1-\varepsilon^2})}} 
\left( \varepsilon \gamma |0\ket + (1-\gamma ) | 2\ket \right) \, ,
 \nonumber \\
|R \ket &= & \frac{-1}{\sqrt{2(1+\sqrt{1-\varepsilon^2})}} 
\left( - \varepsilon \gamma |0\ket + (1+\gamma ) | 2\ket \right) \, .
\eeq
where 
$\gamma = 1/{\sqrt{1-\varepsilon^2}}$, 
$\Delta =  (E_2 - E_0) / 2 $.  
Numerical values for eigenvalues in units of $\omega$ are
\beq
E_0 = 3.0\, , \; \; \; E_2 = 3.751\, , \; \; \; 
\varepsilon = -0.538\, , 
\eeq
and therefore, we find 
\beq
E_{ N} = 2.930\, , \; \; \; E_{R} = 3.821 \, .
\eeq
As expected, the mass of the nucleon is lowered
 and the mass of the Roper is pushed up, where the
%revised..
order of the effect can be estimated as
\beq
\frac{(E_{R} -E_{N}) - (E_{2}-E_{0})}
{E_{2}-E_{0}}  = \frac{ 0.140}{  0.891}=  0.157 \, .  \nonumber
\eeq
We find that the effect on the mass difference is of order  
\( \varepsilon ^{2} \). 
In contrast, the wave functions (orthonormal states) 
are modified substantially:
\beq
|N\ket &=& +1.139 |0\ket + 0.332 |2\ket \, , \nonumber \\
|R\ket &=& +0.332 |0\ket + 1.139 |2\ket \, .
\eeq

Using the wave functions (\ref{NRdiag}), transition amplitudes 
are modified as
\beq
\label{Mdiag}
\bra R | \calM |N\ket 
=
- \sgn(\varepsilon) \frac{1}{1-\varepsilon^2}
\bra 0 | \calM | 2 \ket 
+ \frac{|\varepsilon|}{2(1-\varepsilon^2)}
\left( \bra 0 | \calM | 0\ket 
+ \bra 2 | \calM | 2\ket \right) \;.
\eeq
Before the diagonalization only the first term with the unit 
coefficient ($\varepsilon \to -0$) contribute.  
The new terms, the second and the third terms, have the opposite 
sign to the first term, and therefore, the effect of the 
diagonalization is to reduce the amplitude.  
Substituting numerical numbers for the $\varepsilon$ and matrix elements, 
we obtain 
\beq
A_{1/2}^p &=& 153 - 60 - 57 = 36 \; 
{\rm GeV}^{-1/2}\, , 
\nonumber \\
A_{1/2}^n &=& -102 +40 +38  = -24\; 
{\rm GeV}^{-1/2}\, . 
\label{Adiagonal}
\eeq
We see that the diagonalization affects more on 
transition amplitudes than on masses.  
This is not surprising, since the effect is of 
order $\varepsilon^2$ for masses, while it is of 
order of $\varepsilon^1$ for transitions.  
We note that the modified values (\ref{Adiagonal}) are now rather 
close to those of $d=1$.  
% What is unexpected here is that the change in the 
% amplitude is large such that it turns back close to the original 
% value  of the 
% spherical quark model ($\sim 23$ GeV$^{-1/2}$ and $-15$
% GeV$^{-1/2}$, 
% for the protn and neutron, respectively).  

%------------------------------------------------------------------
\subsection{Relativistic effects}
%------------------------------------------------------------------

The importance of relativistic effects was first emphasized by Kubota 
and Ohta already twenty years ago~\cite{kubota}. 
Further extensive studies were also performed by Capstick and 
Keister~\cite{capstick,capstick2}.
It should be particularly so for transitions where the leading order 
contributions in the long wave length limit are forbidden due to 
some trivial selection rules.  
The M1 transition to the Roper resonance is a typical
example of such; 
the M1 operator with a spin flip leaves the spatial wave 
function unaffected and therefore the matrix element vanishes due to 
the orthogonality between the initial and final spatial wave 
functions in the limit of zero momentum transfer, $k \to 0$.  
Hence the non-zero matrix element at finite $k$ 
is of order $\calO (k)$ where the 
relativistic corrections start to occur.  

So far the fully relativistic treatment has not been achieved yet, 
since it inevitably requires field theoretical method for 
interacting light quarks.   
In literatures, instead, a classical method of $1/m$ 
expansion has been often 
considered~\cite{capstick,ClLi,kubota,capstick2,brodsky}.  
We should note that this method suffers from a fundamental question 
whether $1/m$ expansion converges or not, when $m \sim 300$ MeV and 
$q \sim 500$ MeV.  
Although there is such a difficulty, still we think that the leading 
correction provides us with some flavor how relativistic effects
appear.  

% Here we simply estimate leading corrections in the DOQ model.  
% We perform this for the Roper state, since this is the channel where 
% the disagreement between the theory and experiments is most 
% remarkable.  

The leading order contribution of the electromagnetic hamiltonian
\beq
H_{\gamma} = -e \int d^3x \bar \psi \vec \gamma \psi \scdot \vec A
\;,
\eeq
leads to the standard form of (\ref{Jem}).  
The next to leading order term is of our concern here.  
There are two terms which are so called spin-orbit and non-additive 
terms.  
Here we consider the spin-orbit contributions in order to see the 
importance of relativistic corrections.   
Performing the Foldy-Wouthuysen transformation up to order $1/m^2$, 
we find the relevant terms:
\beq
\label{Hemrel}
H_{\gamma} &\sim& \frac{e}{2m} u_f^\dagger \left( -2 \vec A \cdot \vec p 
- i \vec \sigma \cdot \vec k \times \vec A \right) u_i 
%\nonumber \\
+  \frac{ike}{8m^2} u_f^\dagger \left(
\vec \sigma \cdot \vec k \times \vec A - 2\vec \sigma \cdot 
\vec A \times \vec p \right) u_i \, . 
\eeq
Here $\vec k$ is the photon momentum and $\vec p$ the momentum of the 
initial nucleon (acting on $u_i$).  
In deriving (\ref{Hemrel}), we have used the condition for 
the real photon
$\vec k \cdot \vec A = 0$.

These relativistic corrections are computed using $m \sim 300$ MeV 
in the DOQ model after the diagonalization is performed.  
We find, after including the corrections, the total amplitudes: 
\beq
A^{p}_{1/2} &=& -13 \times 10^{-3}\; {\rm GeV^{-1/2}} \, , \nonumber \\
A^{n}_{1/2} &=& 9 \times 10^{-3} \; {\rm GeV^{-1/2}} \, .
\label{A12relcorr}
\eeq
As in previous works, we find that also in the DOQ model the 
relativistic correction changes the sign of the amplitudes.  
Although the DOQ values in (\ref{A12relcorr}) are larger 
than those of the traditional quark model, they are still too small as 
compared with experimental values.  
We note the similarity in the behaviors of the numbers in the 
conventional and the DOQ models when relativistic effects are
included.  
In both cases, relativistic corrections are very large.  
If so, there is an essential question on the validity of the 
$1/m$ expansion.  
This is a very important problem for the non-relativistic quark model, 
although the model seems to work well for many phenomenological 
aspects.
% ~\cite{shimizu}
.  

%------------------------------------------------------------------
\subsection{Limits in a non-relativistic treatment}
%------------------------------------------------------------------

% In the previous subsection, we have discussed that relativistic 
% corrections have significant effects.  
Here, focusing on the electromagnetic transition of the Roper, we 
demonstrate that within a non-relativistic treatment of a naive quark 
model it is not possible to solve the sign problem of the amplitude.  
Although the statement is rather trivial, it is useful to point it out 
here.  

As emphasized in the preceding discussions, we need to compute both 
the electromagnetic and the strong (pion) couplings when discussing the sign.  
To leading order in the non-relativistic expansion, transition 
operators for these two couplings are given by (\ref{Jem}) and 
(\ref{Hpi}).  
Let us assume, as in the naive quark model,  
that both the nucleon and the Roper is dominated by the 
orbital wave function of $l = 0$: 
$|N\ket \sim |{\rm Roper} \equiv R\ket \sim |[0,1/2]^{1/2}\ket$.  
Then, it is not difficult to show that both the couplings contain the 
spin matrix element $\bra \sigma \ket$:
\beq
\bra N | H_{\pi} | R \ket \cdot \bra R | H_{\gamma} | N \ket 
\sim
\bra N || \sigma  || R \ket \cdot \bra R || \sigma  || N \ket 
\sim 
|\bra N || \sigma  || R \ket |^2
\, .  \nonumber
\eeq
Therefore, it is apparent that within this treatment the sign of the 
amplitude can not be changed.  
We need more delicate treatment for a better understanding of 
transition amplitudes.  

%==================================================================
%SECTION{Summary}
\setcounter{equation}{0}
\renewcommand{\theequation}{\arabic{section}-\arabic{equation}}
\section{Summary}
%==================================================================

In this paper we have 
studied electromagnetic transitions of
excited baryons using a non-relativistic 
quark model with a possibility for excited baryon being deformed.
The main purpose of the present work is to test 
the success of the deformed oscillator quark (DOQ) 
model for the masses of 
flavor $SU(3)$ baryons by examining electromagnetic transitions.  
% There are several possibilities of transitions such as 
% $N + \gamma \to N^{*}$ and $N^{*} + \gamma \to N^{* \prime}$.  
Experimentally, transitions from the ground state have been observed in 
pion photoproductions. 

We have derived all necessary formulae for multipole amplitudes, 
which are transformed into the conventional helicity amplitudes.  
In the comparison of the theoretical amplitudes with experimental data, we have 
paid a special attention to relative signs among amplitudes by 
computing the pion couplings explicitly.  

Our main interest is how electromagnetic amplitudes are influenced 
by the spatial deformation.   
We have seen, however, rather small dependence on deformation, which is 
typically less than 50~\%.  
To this order, there are many other effects which modifies theoretical 
predictions.  
Furthermore, experimental amplitudes usually contain even larger 
ambiguities due to difficulties in the analysis.  
Other types of transitions might be useful to investigate.  
They could be transitions between excited states through photon or 
pion emission.  

Another interest has been paid to the transition to the Roper 
resonance.  
The discrepancy between theoretical predictions and 
experimental data both in magnitude and sign should be considered 
seriously.  
It does not seem that there is a simple explanation so far, 
if we recall the success of the 
magnetic moments for the ground states of the nucleon and delta.  
Further investigations for the Roper resonance will be an interesting 
subject.

\section*{Appendix}

\appendix

%==================================================================
%SECTION{Appendix A}
\section{Helicity formalism}
\setcounter{equation}{0}
\renewcommand{\theequation}{\Alph{section}-\arabic{equation}}
%==================================================================

Empirical amplitudes for the pion photoproduction is presented by the helicity 
coefficients $C^{l\pm}_{\lambda}$ of (\ref{Clmd1}) and (\ref{Clmd2}).  
In order to compute them we explore basic ingredients for the helicity formalism,  
following the classic paper by Jacob and Wick~\cite{JacWick}.   
Let us assume that the relative momentum between the initial 
photon and the 
nucleon is along the $z$-axis with their helicities being 
$\lambda_{\gamma}$ and $\lambda_{i}$.  
In the center of mass system, the total helicity is given by 
$\lambda = \lambda_{\gamma} - \lambda_{i}$.  
This is equivalent to the $z$ component of the angular momentum in the 
spherical basis.    
Let us write the initial state as
\beq
\label{k00lam}
|k,\theta \phi, \lambda\ket = |k,00, \lambda\ket\, ,
\eeq
where $\theta \phi= 00$ are the polar angles of the momentum 
$\vec k$.   
Similarly we write  the final state for the pion and nucleon 
system as
\beq
\label{qabmu}
|q,\theta \phi, \mu\ket \, , 
\eeq
where $\theta \phi  $ are the scattering angles of the pion, and  
$\mu$ is the helicity of the final state.  

The helicity states (\ref{k00lam}) and (\ref{qabmu}) can be 
related to the spherical states through
\beq
\label{p_expand}
|p, \theta \phi, \lambda\ket  = \sum_{jm}
\sqrt{\frac{2j+1}{4\pi}} D^{j}_{m\lambda}(\phi, \theta, -\phi)
|p, jm, \lambda\ket \, , 
\eeq
and the inverse relation
\beq
\label{jm_expand}
|p, jm, \lambda\ket = 
\sqrt{\frac{2j+1}{4\pi}}
\int d\Omega \, D^{j*}_{m\lambda}(\phi, \theta, -\phi) 
|p, \theta \phi, \lambda\ket \, .
\eeq
In these equations, $D^j_{mn}(\alpha, \beta, \gamma)$ are the 
Wigner's $D$-function~\cite{VMK88}.  
For rotation of a vector, the angle $\gamma$ is 
redundant and can be fixed as $\alpha = -\gamma \equiv \phi$.  
The integral measure is then given by 
$d\Omega = d(\cos\theta) d\phi$, and so
$\int d\Omega = 4\pi$.  

Applying (\ref{p_expand}) to (\ref{k00lam}) and (\ref{qabmu}), we can 
rewrite the helicity amplitude (\ref{Aexpand})
\beq
A_{\mu\lambda}(\theta\phi) &\equiv&
\bra q, \theta \phi, \mu | \calA | k, 00, \lambda, \ket 
\nonumber \\
&=&
\sum_{jj'm} \sqrt{\frac{2j'+1}{4\pi}} \sqrt{\frac{2j+1}{4\pi}}
D^{j*}_{m\mu}(\phi, \theta, -\phi) 
\bra q, jm, \mu|\calA | k, j'\lambda ,\lambda\ket \, .
\eeq
On account of the conservation of angular momentum, 
$j' = j$ and $m = \lambda$, we obtain
\beq
\bra q, \theta \phi, \mu | \calA | k, 00, \lambda\ket 
=
\frac{1}{4\pi} \sum_{j} (2j+1) 
D^{j*}_{\lambda \mu}(\phi, \theta, -\phi) 
\bra q, j\lambda, \mu|\calA | k, j\lambda, \lambda \ket \, , 
\eeq
which corresponds to (\ref{Aexpand}).  

It is instructive to re-express these relations using spherical 
harmonics.  
For this we consider 
\beq
|k,00, \lambda = 1/2 \ket &\sim& e^{i\vec k \cdot \vec x} \, 
\chi_{1/2}
\; = \; 
4\pi \sum_{l_{\pi} m_{\pi}} i^{l _{\pi}} \, 
Y_{l_{\pi}m_{\pi}}(\hat x) Y_{l_{\pi}m_{\pi}}^*(\hat k)\, 
j_{l_{\pi}}(kr) \, \chi_{1/2}\, , 
\eeq
where $\chi_{1/2}$ is the spin up state for the nucleon.  
Using 
$Y_{l_{\pi}m_{\pi}}^*(\hat k = \hat z) =\sqrt{(2l_{\pi}+1)/4\pi} \, 
\delta_{m_{\pi}0}$ and 
recoupling the angular momentum, we find
\beq
\label{p00exp+}
|k,00, 1/2\ket \sim
\sqrt{2 \pi} \sum_j \sqrt{2j+1}
\left( i^{j-1/2} \calY^{j-1/2}_{j\, 1/2} 
- i^{j+1/2} \calY^{j+1/2}_{j\, 1/2} \right) \, ,
\eeq
where $\calY^{l_{\pi}}_{j\, m}$ is the spinor harmonics
including the spherical Bessel functions $j_{l_{\pi}}(kr)$:
\beq
\label{p00exp-}
\calY^{l_{\pi}}_{j\, m} = %j_{l_{\pi}}(kr) 
\sum_{\mu} 
(l_{\pi}\, m \! - \! \mu \, 1/2 \, \mu \, | j \, m) 
\mathscr{Y}_{l_{\pi}\, m \! - \! \mu}\, \chi_{\mu} \, , 
\eeq
where $\mathscr{Y}_{lm} = Y_{lm}j_{l}(kr)$.  
Similarly, we find 
\beq
|k,00, -1/2\ket \sim
\sqrt{2 \pi} \sum_j \sqrt{2j+1}
\left( i^{j-1/2} \calY^{j-1/2}_{j\, 1/2} 
+ i^{j+1/2} \calY^{j+1/2}_{j\, 1/2} \right) \, .
\eeq

The states with a definite angular momentum can be obtained 
by applying the projection operator: 
\beq
\label{pjm_conc}
|k, jm, \pm 1/2 \ket &\sim&
\sqrt{\frac{2j+1}{4\pi}} \int d\Omega \, 
D^{j*}_{m\, \pm 1/2} (\phi, \theta, -\phi) 
R(\phi, \theta, -\phi)  \, |k, 00, \pm 1/2\ket 
\nonumber \\
&=& 
\frac{4 \pi}{\sqrt{2}}
\left( i^{j-1/2} \calY^{j-1/2}_{j\, 1/2} 
\mp i^{j+1/2} \calY^{j+1/2}_{j\, 1/2} \right) \, . 
\eeq
Here 
$R(\phi, \theta, -\phi) = 
\exp(-i\alpha J_z)\exp(-i\beta J_y)\exp(-i\gamma J_z)$ 
is the rotation operator, and in deriving the second equation we have 
used the definition of the $D$-functions.
Obviously, in these states (\ref{pjm_conc}) 
parities are mixed.   
States with a definite parity can be obtained by
\beq
\label{pjmpm}
|p, jm, \pm\ket = \frac{1}{\sqrt{2}}
\left( |p, jm, + 1/2\ket \pm |p, jm, - 1/2\ket \right)
= \pm 4 \pi \, i^{j \mp 1/2} \, \calY^{j \mp 1/2}_{jm} \, .
\eeq
These are the eigenstates of parity 
$P = \eta (-)^{l _{\pi}} = \eta (-)^{j\mp1/2}$, 
where $\eta$ is the intrinsic parity of the system.

%==================================================================
%SECTION{Appendix B}
\section{The pion matrix element and the sign $\epsilon$}  
\setcounter{equation}{0}
\renewcommand{\theequation}{\Alph{section}-\arabic{equation}}
%==================================================================

Let us start with the relation (\ref{Almd}):
\beq
\label{Almdmod}
A_\lambda^{j^{P}} &=& 
-i \left( \frac{1}{(2j+1)\pi} \frac{k}{q} \frac{M}{M^*} 
\frac{\Gamma_\pi}{\Gamma^2} \right)^{-1/2}\, 
C^{l_{\pi}\pm}_{\lambda}(M^{*}) C_{\pi N}\\
&=& 
\;  ( \; \cdots \; ) \; 
\bra q, j\lambda, \pm |H_{\pi}|N^*(j\lambda)\ket 
\, 
G_{j}^{N^*}
\, 
\bra N^*(j\lambda)|H_{\gamma}|k, j\lambda, \lambda\ket C_{\pi N}
\, .
\eeq
We need to compute the pion matrix element 
$\bra q, j\lambda, \pm |H_{\pi}|N^*(j\lambda)\ket$ 
in order to account for the sign of the whole amplitude.  

To do so, it is convenient to have an explicit form for the final 
state as given in (\ref{pjmpm}):
\begin{equation}
\begin{split}
    \label{calYf}
    \bra q, j\lambda, \pm |
    =& 
    \left( \pm i^{l_{\pi}} |\calY^{l_{\pi}}_{j\lambda}\ket \right)^\dagger 
    =
    \left( \pm  i^{l _{\pi}} 
    \sum_m 
    %(l_{\pi}\, m\, 1/2\, \lambda \! - \! m|j\lambda) 
    \CG{l_{\pi}}{m}{1/2}{\lambda}{\! - \! m}{j}
    \mathscr{Y}_{l_{\pi}m}|\chi_{\lambda-m}\ket \right)^\dagger 
    \\
    =&
    \pm (-i)^{l_{\pi}} 
    \sum_m 
    % (l_{\pi}\, -m\, 1/2\, \lambda \! + \! m|j\lambda)
    \CG{l_{\pi}}{\!-\!m}{1/2}{\lambda\! + \! m}{j}{\lambda}
    \, (-)^m \, 
    \bra \chi_{\lambda+m}|\mathscr{Y}_{l_{\pi}m} \, .
\end{split}
\end{equation}
In this equation, $l_{\pi} \equiv j \mp 1/2$ 
is the orbital angular momentum for the pion relative to the nucleon.  
The interaction hamiltonian for the pion-quark coupling is given by 
\beq
H_{\pi} = - \frac{g}{2m} \vec \sigma \scdot \vec \nabla_{\pi} \, ,
\label{Hpi}
\eeq
where the derivative $\nabla_{\pi}$ operates to the pion wave 
function.  
Sandwiching this hamiltonian with (\ref{calYf}) and the resonance 
state $|N^*(j\lambda)\ket$, we find after some algebra 
\beq
\label{H_pi1}
\bra q, j\lambda, \pm | H_{\pi} | N^*(j\lambda)\ket  & & \nonumber \\
& & \hspace{-4cm} = \; 
\mp 4 \pi \frac{gq}{2m} i^{l_{\pi}}
\left( \sqrt{\frac{l_{\pi}+1}{2l_{\pi}+1}} 
\bra 1/2 || [\mathscr{Y}_{l_{\pi}+1} \sigma]^{l_{\pi}} 
                        || j \ket 
   + \sqrt{\frac{l_{\pi}}{2l_{\pi}+1}} 
\bra 1/2 || [\mathscr{Y}_{l_{\pi}-1} \sigma ]^{l_{\pi}} 
                        || j \ket \right) \, .
\eeq
In fact, one of the two terms is relevant depending on the 
orbital angular momentum $l_q \equiv l$ of an excited quark in $N^*$.  
Due to the pseudoscalar nature of the pion-quark coupling, it follows 
that $l = l_{\pi} \pm 1$.    
When $j = l -1/2$ or $j = l -3/2$, the first term of (\ref{H_pi1}) 
survives, 
while when $j = l +3/2$ or $j = l + 1/2$, the second term does.  

For later convenience when performing actual computations, we 
summarize relevant pion matrix elements according to 
$j = l+3/2,\; l+1/2,\; l-1/2,\; l-3/2$: 

\begin{enumerate}
    \item  When $j = l +3/2,\;  l+1/2 \; \; 
%           \stackrel{l_{\pi} = l+1}{\longrightarrow} 
    \xrightarrow{l_{\pi} = l+1}
    \; \; j = l_{\pi} \pm 1/2$ \, , 
    \beq
    \bra q, j\lambda\, \pm | H_{\pi} | N^*(j\lambda)\ket
    \; = \; 
    \mp 4 \pi \frac{gq}{2m} i^{l+1} \sqrt{\frac{l+1}{2l+3}} 
    \bra 1/2 || [\mathscr{Y}_l \sigma]^{l+1}  || j \ket \, .
    \eeq

    \item  When $j = l -1/2 , l - 3/2\; \; 
%            \stackrel{l_{\pi} = l-1}{\longrightarrow} 
    \xrightarrow{l_{\pi}= l-1} 
           \; \; j = l_{\pi} \pm 1/2$\, , 
           \beq
        \bra q, j\lambda\, \pm | H_{\pi} | N^*(j\lambda)\ket
        \; = \;  
        \pm 4 \pi \frac{gq}{2m} i^{l+1} \sqrt{\frac{l}{2l-1}} 
        \bra 1/2 || [\mathscr{Y}_l \sigma]^{l-1} || j \ket \, .
        \eeq

\end{enumerate}
Writing the pion matrix element as 
\beq
\bra q, j\lambda\, \pm | H_{\pi} | N^*(j\lambda)\ket
\equiv
-i (-i)^{l} \; \epsilon \; 
| \bra q, j\lambda\, \pm | H_{\pi} | N^*(j\lambda)\ket | \, , 
\eeq
the sign $\epsilon$ is given as summarized in Table~\ref{signpi}.  
In the table, $\eta$ is the sign of the radial integral
$\sum_{n} F^{N\sigma}_{00nl} (00|j_{l}|nl)$ .  
The notations of this equation is defined in (\ref{def_radint}).

\begin{table}[tbh]
        \centering
        \footnotesize
        \caption{\small 
        The sign $\epsilon$ extracted from the pion matrix 
        elements.
        \label{signpi}
        }
        \vspace{0.5cm}
        \begin{tabular}{ c c c | c c c}
\hline
\multicolumn{3}{c|}{Nucleon} & \multicolumn{3}{c}{Delta}\\
 & $j$ & $\epsilon$ &  & $j$ & $\epsilon$\\
\hline
$S = 1/2$ & $l+1/2$ & $-\eta_{S\; {\rm or}\; \lambda}$ 
                   & $S = 1/2$ & $l+1/2$ & $+\eta_{\lambda}$\\
          & $l-1/2$ & $+\eta_{S\; {\rm or}\; \lambda}$ &
                               & $l-1/2$ & $-\eta_{\lambda}$\\
$S = 3/2$ & $l+3/2$ & $+\eta_{\lambda}$ 
                   & $S = 3/2$ & $l+3/2$ & $+\eta_{S}$\\
          & $l+1/2$ & $-\eta_{\lambda}$ & 
                               & $l+1/2$ & $-\eta_{S}$\\
          & $l-1/2$ & $-\eta_{\lambda}$ &
                               & $l-1/2$ & $-\eta_{S}$\\
          & $l-3/2$ & $+\eta_{\lambda}$ &
                               & $l-3/2$ & $+\eta_{S}$\\
\hline
          \multicolumn{6}{c}
          {$\eta_{\sigma} \equiv \sgn \left[
          \sum_{n} F^{N\sigma}_{00nl} (00|j_{l}|nl)
          \right]$} \\
\end{tabular}
\end{table}

%==================================================================
%SECTION{Appendix C}
\section{Computation of multipole amplitudes}
\setcounter{equation}{0}
\renewcommand{\theequation}{\Alph{section}-\arabic{equation}}
%==================================================================

Using the wave functions for $N^*$ as given in 
(\ref{None}) -- (\ref{Dtwo}) 
one can compute multipole amplitudes in a straightforward manner.  
For a given excited baryon, there are three possible amplitudes:
$\calM (M\, l+1)$, $\calM (M\, l-1)$ and $\calM (E\, l)$.  
To illustrate how actual computations can be performed, 
we consider one example for 
$\calM(M\, l+1)$ for the transition
$ |N(\text{ground state})\ket \to |N^{*}; [l_{S},\, 1/2]^j \ket$.  

In the following expressions, the factor 3 of (3-10) is 
not included, although it is in our numerical results.  
Replacing $l \to l+1$ in (\ref{Mlgeneral}), 
only the second term survives due to the 
matching of orbital angular momentum.  
Since the initial state for the ground state nucleon 
is spherical, $l = 0$, the orbital 
angular momentum of 
the absorbed photon and that of the excited baryon must be the same.  
Thus we have
\beq
\calM(M\, l+1) &=&    
- \frac{k}{4m} 
\sqrt{ \frac{l+2}{2l+3} } 
\left(
\bra [\Psi^{NS}_{l} \chi^\rho ]^j || [\mathscr{Y}_l \sigma]^l ||
\Psi^{0S}_{l=0} \chi^\rho \ket \tau_\mu(\rho) 
\right. \nonumber \\
\label{calM1}
& & \hspace{2.5cm} + \; \left.
\bra  [\Psi^{NS}_{l} \chi^\lambda ]^j || [\mathscr{Y}_l \sigma]^l ||
\Psi^{0S}_{l=0} \chi^\lambda \ket \tau_\mu(\lambda) 
\right) \, . 
\eeq
Here we have included the isospin matrix elements which are 
defined by
\beq
\label{taurho}
\tau_\mu(\rho) = \bra 
\phi^\rho | \tau_\mu |\phi^\rho \ket \, , \;\;\;
\tau_\mu(\lambda) = \bra \phi^\lambda | \tau_\mu |\phi^\lambda \ket 
\; ,
\eeq
and the radial integral $\int r^2 dr$ is implicit.   
Now the reduced matrix elements can be computed in a straightforward 
manner using the decomposition theorems, and we find the result
\beq
\calM(M\; l+1) &=& - \frac{k}{4m}
\sqrt{\frac{l+2}{2l+3}} \; \hat j \; \widehat{l+1} 
%\left\{
%\begin{array}{ccc}
 \begin{Bmatrix}
        1/2 & j & l  \\
        l+1 & 1 & 1/2
\end{Bmatrix} 
%\end{array}\right\} 
\nonumber \\
\label{Ml+1_N1*}
& \times &
\frac{1}{4\pi} \sum_n F^{NS}_{nl00} \; (nl| j_l| 00) \; \sqrt{6} \; 
(\tau_\mu(\rho) - \frac{1}{3} \tau_\mu(\lambda)) \; ,
\eeq
where \( \hat{l} =\sqrt{2l+1} \).
Here we have introduced the notation for the radial matrix elements: 
\beq
\label{def_radint}
(nl| j_l| 00) &=& \int r^2 dr R_{nl} j_l R_{00} \; ,
\eeq
where $R_{nl}$ are the radial functions of the harmonic oscillator.  

All other matrix elements can be computed in similar ways.  
Here we summarize the results for all amplitudes.  
Let us introduce the following 
notation for the sum of the radial matrix elements:
\beq
\label{Psum}
P^{N\sigma}_l &=& 
\frac{1}{\sqrt{4\pi}} \sum_n F^{N\sigma}_{00nl} \; (nl| j_l| 00) \; , \\
\label{Qsum}
Q^{N\sigma}_l &=& 
\frac{1}{\sqrt{4\pi}} \sum_n F^{N\sigma}_{00nl} \; (nl^\pr| j_l| 00) 
\; , 
\eeq
and the combinations of isospin matrix elements
\beq
T_{\pm 1} &=& \tau_\mu(\rho) \pm \tau_\mu(\lambda) \nonumber \; ,
\\
T_{\pm 1/3} &=& \tau_\mu(\rho) \pm \frac{1}{3}\tau_\mu(\lambda) \; .
\nonumber
\eeq
We checked the behavior of the coefficients \( F \) numerically and 
found that they converge to zero quickly as \( n  \) is increased.  
In the present work, we take the sum over nine terms in
\eqref{Psum} and \eqref{Qsum}
to obtain the accuracy of more than five digits.  

The matrix elements for various excited states $N^{*}$ are 
summarized as follows:  

\begin{description}

%======== case N1
\item[N1]  $N^* = |[l_{S},\, 1/2]^j \ket$:
%\calM(M\; l+1)
\beq
\hspace{-1cm}
\label{Ml+1_N1}
\calM(M\; l+1) =  - \frac{k}{4m}
\sqrt{\frac{l+2}{2l+3}} \; \hat j \; \widehat{l+1}
% \left\{ \footnotesize 
% \begin{array}{ccc}
 \begin{Bmatrix}
        1/2 & j & l  \\
        l+1 & 1 & 1/2
\end{Bmatrix} 
% \end{array}
% \right\} 
\times 
P^{NS}_l \; \sqrt{6} \; T_{-1/3}
\eeq
%\calM(M\; l-1)
\beq
\label{Ml-1_N1}
\hspace{-1cm}
\calM(M\; l-1) =  \frac{k}{4m}
\sqrt{\frac{l-1}{2l-1}} \; \hat j \; \widehat{l-1}
% \left\{ \footnotesize 
% \begin{array}{ccc}
 \begin{Bmatrix}
        1/2 & j & l  \\
        l-1 & 1 & 1/2
\end{Bmatrix} 
% \end{array}
% \right\} 
P^{NS}_l \; \sqrt{6} \; T_{-1/3}
\eeq
%\calM(El)
\beq
\hspace{-1cm}
\calM(El)  =  
- \frac{\sqrt{l(l+1)}}{2m}(-1)^l \; 
\hat j
\; Q^{NS}_l \; T_{+1} 
\label{El_N1}
+ \frac{k}{4m} \; \hat j\;  \hat l\; 
% \left\{ \footnotesize 
% \begin{array}{ccc}
 \begin{Bmatrix}
        1/2 & j & l  \\
        l & 1 & 1/2
\end{Bmatrix} 
% \end{array}
% \right\}
P^{NS}_l \; \sqrt{6} \; T_{-1/3}
\eeq

%======== case N2
\item[N2]  
$N^{*} = |[l_{MS},\, 1/2]^j \ket$:

This case is obtained simply by performing the following replacements 
in the results of (N1).  
\begin{itemize}
        \item  Multiply $1/\sqrt{2}$ as an overall factor. 

        \item  Change the sign of one of the isospin matrices:
        $+ \tau_\mu(\lambda) \to - \tau_\mu(\lambda)$. 

        \item  Replace the superscript of the coefficient $F$:
        $F^{NS}_{00nl} \to F^{N\lambda}_{00nl}$ and hence 
        $P^{NS}_l \to P^{N\lambda}$ and 
        $Q^{NS}_l \to Q^{N\lambda}$.  
\end{itemize}

%======== case N3
\item[N3]   
$N^{*} =  |[l_{MS},\, 3/2]^j \ket $:
%\calM(M\; l+1)
\beq
\label{Ml+1_N3}
\hspace{-1cm}
\calM(M\; l+1) = - \frac{k}{4m}
\sqrt{\frac{l+2}{2l+3}} \; \hat j \; \widehat{l+1}
% \left\{ \footnotesize 
% \begin{array}{ccc}
 \begin{Bmatrix}
        3/2 & j & l  \\
        l+1 & 1 & 1/2
\end{Bmatrix} 
% \end{array}
% \right\}  
P^{N\lambda}_l \; 
\frac{4}{\sqrt{3}} \; \tau_\mu(\lambda) 
\eeq
%\calM(M\; l-1)
\beq
\label{Ml-1_N3}
\hspace{-1cm}
\calM(M\; l-1) = \frac{k}{4m}
\sqrt{\frac{l-1}{2l-1}} (-1)^{j+3/2} \; \hat j \; \widehat{l-1}
% \left\{ \footnotesize 
% \begin{array}{ccc}
 \begin{Bmatrix}
        3/2 & j & l  \\
        l-1 & 1 & 1/2
\end{Bmatrix} 
% \end{array}
% \right\} 
 P^{N\lambda}_l \; 
\frac{4}{\sqrt{3}} \; \tau_\mu(\lambda) 
\eeq
%\calM(El_Na)
\beq
\label{El_N3}
\hspace{-1cm}
\calM(El) = 
- \frac{k}{4m} \; \hat j\;  \hat l\; 
% \left\{ \footnotesize 
% \begin{array}{ccc}
 \begin{Bmatrix}
        3/2 & j & l  \\
        l & 1 & 1/2
\end{Bmatrix} 
% \end{array}
% \right\}  
 P^{N\lambda}_l \; 
\frac{4}{\sqrt{3}} \; \tau_\mu(\lambda) 
\eeq

\end{description}

The decays of delta excited states $\Delta$1 and $\Delta$2 
are obtained by the following manipulations.

\begin{description}

%======== case D1
\item[${\bf \Delta}$1]  
$\Delta^* = |[l_{MS},\, 1/2]^j \ket$:
In the results of {\bf N1}
\begin{itemize}
        \item  Pick up the $\tau_{\mu}(\lambda)$-term and replace 
        $\tau_{\mu} 
        \to \bra \phi^{\lambda} | \tau_{\mu} | \phi^{S} \ket$ .

        \item  Replace $F^{NS}_{00nl}$ by $F^{N\lambda}_{00nl}$.  
\end{itemize}

%======== case D2
\item[${\bf \Delta}$2]  $\Delta^* =  |[l_{S},\, 3/2]^j \ket$:
In the results of {\bf N3}
\begin{itemize}
        \item  Replace $\tau_{\mu} 
        \to \bra \phi^{\lambda} | \tau_{\mu} | \phi^{S} \ket$ .
        
        \item  Replace $F^{NS}_{00nl}$ by $F^{N\lambda}_{00nl}$ .
        
        \item Multiply $\sqrt{2}$ as an overall factor.  
\end{itemize}

\end{description}

For practical purposes, it is convenient to write the spin 
of excited baryons $j$ by $l$ ($j = l \pm 1/2$ or $l \pm 3/2$), so
that 
the 6-$j$ symbols are computed explicitly.  
Results are: 
%
%%%%%%%%%%%%%

\begin{description}
%======== case N1.a
\item[N1.a]  $N^* =  |[l_{S},\, 1/2]^{l+1/2} \ket$, 
allowed transitions: $M \; l+1$ and $E\; l$, 
\beq
\label{M_N1a}
\calM(M \; l+1) = - \frac{k}{4m} \sqrt{2(l+2)}  \; P^{NS}_l  
\; T_{-1/3}
\eeq
\beq
\label{E_N1a}
\calM(E \; l) = \frac{\sqrt{2l}(l+1)}{2m}  \; Q^{NS}_l  
\; T_{+1}
+ \frac{k \sqrt{2l}}{4m} \; P^{NS}_l  \; T_{-1/3}
\eeq

%======== case N1.b
\item[N1.b]  $N^* =  |[l_{S},\, 1/2]^{l-1/2} \ket$, 
allowed transitions: $M  \;  l-1$ and $E \; l$, 
\beq
\label{M_N1b}
\calM(M  \;  l-1) = \frac{k}{4m} \sqrt{2(l-1)} \; P^{NS}_l 
\; T_{-1/3}
\eeq
\beq
\label{E_N1b}
\calM(E  \;  l) = - \frac{\sqrt{2(l+1)}l}{2m}  \; Q^{NS}_l  
\; T_{+1}
+ \frac{k \sqrt{2(l+1)}}{4m} \; P^{NS}_l  \; T_{-1/3}
\eeq

%======== case N2
\item[N2]  
The results for this case, where 
$N^* = |[l_{S},\, 1/2]^{l \pm 1/2} \ket$
are obtained by making the replacements as described before.

%======== case N3.a
\item[N3.a]  $N^* = |[l_{MS},\, 3/2]^{l+3/2} \ket$, 
allowed transitions: $M  \;  l+1$, 
\beq
\label{M_N3a}
\calM(M \; l+1) = 
\frac{k}{2m} \sqrt{\frac{2}{3}} \frac{l+2}{\sqrt{2l+3}}
 \; P^{N\lambda}_l  \; \tau_\mu(\lambda)
\eeq

%======== case N3.b
\item[N3.b]  $N^* =  |[l_{MS},\, 3/2]^{l+1/2} \ket$, 
allowed transitions: $M  \;  l+1$ and $E \; l$, 
\beq
\label{M_N3b}
\calM(M  \;  l+1) = 
- \frac{k}{3m} \sqrt{\frac{l(l+2)}{2(2l+3)}} 
 \; P^{N\lambda}_l  \; \tau_\mu(\lambda)
\eeq
\beq
\label{E_N3b}
\calM(E \; l) = 
- \frac{k}{3m} \sqrt{\frac{2l+3}{2}} 
 \; P^{N\lambda}_l  \; \tau_\mu(\lambda)
\eeq

%======== case N3.c
\item[N3.c]  $N^* =  |[l_{MS},\, 3/2]^{l-1/2} \ket$, 
allowed transitions: $M  \;  l-1$ and $E \; l$, 
\beq
\label{M_N3c}
\calM(M  \;  l-1) = 
- \frac{k}{3m} \sqrt{\frac{(l-1)(l+1)}{2(2l-1)}} 
 \; P^{N\lambda}_l  \; \tau_\mu(\lambda)
\eeq
\beq
\label{E_N3c}
\calM(E \; l) = 
 \frac{k}{3m} \sqrt{\frac{2l-1}{2}} 
 \; P^{N\lambda}_l  \; \tau_\mu(\lambda)
\eeq

%======== case N3.d
\item[N3.d]  $N^* = |[l_{MS},\, 3/2]^{l-3/2} \ket$, 
allowed transitions: $M  \;  l-1$, 
\beq
\label{M_N3d}
\calM(M  \;  l-1) = 
\frac{k}{2m} \sqrt{\frac{2}{3}} \frac{l-1}{\sqrt{2l-1}}
 \; P^{N\lambda}_l  \; \tau_\mu(\lambda)
\eeq
\end{description}
Similarly, we can derive the formulae for deltas.

\end{document}